\documentclass[prd,twocolumn,showpacs,nofootinbib,preprintnumbers,amsmath,amssymb,aps,longbibliography,superscriptaddress,floatfix]{revtex4-2}

\usepackage{graphicx,subfigure,epsfig}
\usepackage{dcolumn}
\usepackage{amssymb}
\usepackage{times}
\usepackage{amsmath}
\usepackage{amsfonts}
\usepackage{mathrsfs}
\usepackage{setspace}
\usepackage{latexsym}
\usepackage{bbm}
\usepackage{float}
\usepackage{flafter}
\usepackage{bm}
\usepackage{epstopdf}
\usepackage{color}
\usepackage{multirow}
\usepackage{physics} 
\usepackage[export]{adjustbox}

\usepackage{braket}
\usepackage{array}

\usepackage{tikz}                  

\usepackage{dsfont}

\usepackage{xr}
\usepackage[pdftex,breaklinks,bookmarks=false,colorlinks,linkcolor=blue,citecolor=blue,urlcolor=blue,hypertexnames=false]{hyperref}

\usepackage[normalem]{ulem}

\newcommand{\vmin}{v_{\rm min}}
\newcommand{\vesc}{v_{\rm esc}}
\newcommand{\vmax}{v_{\rm max}}
\newcommand{\kms}{\,{\rm km/s}}
\newcommand{\Op}[1]{\mathcal{O}_{#1}}

\graphicspath{{figs/}}

\begin{document}

\preprint{FERMILAB-PUB-26-0767-T}

\title{Simulation-Informed High-Velocity Dark Matter\\
in the Effective Theory of Direct Detection}%

\author{Xiuyuan Zhang}
\affiliation{Department of Physics and Kavli Institute for Astrophysics and
Space Research, Massachusetts Institute of Technology, Cambridge, MA 02139, USA}

\author{Gonzalo Herrera}
\affiliation{Department of Physics and Kavli Institute for Astrophysics and
Space Research, Massachusetts Institute of Technology, Cambridge, MA 02139, USA}
\affiliation{Laboratory for Particle Physics and Cosmology, Harvard University, Cambridge, MA 02138, USA}

\author{Andreas Thoyas}
\affiliation{Department of Physics, Northeastern University, Boston, MA 02115, USA}

\author{Lina Necib}
\affiliation{Department of Physics and Kavli Institute for Astrophysics and
Space Research, Massachusetts Institute of Technology, Cambridge, MA 02139, USA}

\date{\today}

\begin{abstract}
 The interpretation and reach of direct detection (DD) experiments depend sensitively on the local dark matter (DM) velocity distribution, particularly its high-velocity tail. 
 This tail is crucial for elastic scattering at low DM masses or large momentum transfers, and for inelastic upscattering, where additional kinetic energy is needed to overcome the mass splitting.
In practice, however, phenomenological studies typically rely either on the Standard Halo Model (SHM) or, when an enhanced high-velocity tail contributes to the scattering rate, on the single simulation that includes a large Magellanic Cloud (LMC) analogue, Au-25, whose debris reaches $\simeq 950\kms$.
In this work, we instead build on a previously presented empirically-motivated reconstruction method, to  
construct the local DM velocity distribution, using stellar kinematics from the Gaia space telescope.
We combine this reconstructed Milky Way component with the best LMC analog in the FIRE-2 simulations, and find no local DM faster than $\simeq 900\kms$. 
We compare the resulting distribution with the two commonly used benchmarks, the SHM and Au-25.
We then study how these different velocity distributions affect the interpretation of DD searches across all operators of the non-relativistic effective field theory of DM--nucleon interactions, for both elastic and inelastic DM.
As an illustrative application, we consider the high-energy event observed in the LUX-ZEPLIN extended search. 
Requiring the event to remain compatible with the absence of signal in the surrounding energy sidebands, we observe the inferred inelastic DM mass splitting can be shifted by up to $\sim 130$~keV depending on the assumed high-velocity tail.
Higher-resolution simulations of the high-velocity DM population reaching the detector will therefore be crucial for interpreting a future signal and setting robust limits on DM interactions.
\end{abstract}

\maketitle


\section{Introduction}
The local dark matter (DM) velocity distribution is an important input to the DM interaction rate at direct detection experiments. For decades it was supplied by the Standard Halo Model (SHM), an isotropic Maxwellian truncated at the escape speed~\citep{1986PhRvD..33.3495D,1996APh.....6...87L}, adopted largely for analytic convenience rather than as a faithful description of a halo assembled hierarchically. DM-only \(N\)-body simulations of Milky-Way-mass halos such as Via Lactea and Aquarius first demonstrated that the distribution at the Solar radius can depart appreciably from a Maxwellian, both near its peak and in its high-speed tail, and can retain features associated with individual accretion events~\citep{Diemand:2008in,Springel:2008cc,Vogelsberger:2008qb,Kuhlen:2009vh,Lisanti:2010qx,Mao:2012hf}. Hydrodynamic simulations that include a baryonic disk (e.g.\ Eris~\cite{2011ApJ...742...76G}, EAGLE/APOSTLE~\cite{Schaye_2014, 2015MNRAS.450.1937C, McAlpine_2016, 2016MNRAS.457..844F, Sawala_2016, 2017arXiv170609899T}, FIRE~\cite{Hopkins_2018, Wetzel_2023, wetzel2025secondpublicdatarelease}, and Auriga~\cite{Grand_2017, 2024MNRAS.532.1814G}) subsequently showed that the deeper central potential tends to make the bulk velocity distribution more nearly Maxwellian~\citep{Ling:2009eh,Kuhlen:2013tra,Bozorgnia:2016ogo,Kelso:2016qqj,Sloane:2016kyi,Bozorgnia:2017brl,Hopkins_2018}, while still preserving kinematic substructure inherited from the Galaxy's merger history~\citep{Necib:2018iwb,Evans:2018bqy,Buch:2019aiw}.
The arrival of \textit{Gaia} suggested a complementary, observational route. Because accreted stars and DM originate in the same disrupted satellites and experience the same Galactic potential, the kinematics of the accreted stellar halo can carry information about the corresponding DM components~\citep{Herzog-Arbeitman:2017fte,Necib:2018igl,Evans:2018bqy,OHare:2018trr}. Earlier studies used this connection to motivate stellar proxies or phenomenological modifications of the SHM, but a systematic reconstruction of the local DM velocity distribution from stellar data requires calibrating the mapping between the stellar and DM debris. In recent work, \citet{zhang2026setnightfirebuilding} developed and calibrated such a reconstruction framework using FIRE simulations.
What such reconstructions do not yet settle is the extreme tail; the bulk of the local distribution is sampled by large numbers of particles in simulations and is increasingly constrained by stellar kinematics, whereas the fastest particles in the Solar neighborhood are rare and therefore poorly resolved. The Large Magellanic Cloud (LMC) is particularly important in this regime. Its recent passage can perturb the Milky Way (MW) halo and bring in an additional population of high-speed DM, and simulations containing LMC analogues find that these effects can populate the largest heliocentric velocities~\citep{Besla:2019xbx,2022MNRAS.513L..46D,SmithOrlik:2023pbo}. Much of the recent phenomenology of such an enhanced tail has relied on a single Auriga analogue~\cite{Grand_2017, 2024MNRAS.532.1814G}, Au-25, whose LMC-associated particles extend to unusually high speeds. The extent of this tail, however, is sensitive to the particular realization of the MW--LMC system and to the numerical resolution of the simulation.

On the particle-physics side, the non-relativistic effective field theory (NREFT) of DM--nucleon interactions~\citep{Fan:2010gt,Fitzpatrick:2012ix,Anand:2013yka} provides a general framework for describing direct-detection signals. It replaces the single spin-independent interaction by fourteen operators built from the momentum transfer $q$, the transverse velocity $v^{\perp}$, and the DM and nucleon spins. Operators carrying explicit powers of $q$ or $v^{\perp}$ preferentially weight different regions of the velocity distribution, and can therefore be especially sensitive to its high-velocity population. The astrophysical and particle-physics inputs are thus not separable: the assumed interaction determines which part of $f(\mathbf{v})$ an experiment probes.

This coupling becomes strongest when the minimum speed required to produce a signal, $\vmin$, approaches the largest speeds available in the laboratory frame. The rate is then dominated by the high-velocity tail, making uncertainties in this population especially consequential. For light DM scattering elastically, $\vmin \propto 1/\mu_{\chi T}$ at fixed recoil energy, where $\mu_{\chi T}$ is the reduced mass of the DM $m_\chi$ and that of the target $m_T$, so only the fastest particles can exceed the detector threshold. Large recoil energies likewise select the high-speed population. This dependence is particularly explicit for inelastic upscattering into a
state heavier by a mass splitting $\delta$~\citep{TuckerSmith:2001hy}.
The minimum speed required to produce a nuclear recoil of energy $E_R$ is
then
\begin{equation}
\vmin(E_R) = \frac{1}{\sqrt{2 m_T E_R}}
\left|\frac{m_T E_R}{\mu_{\chi T}} + \delta\right| ,
\label{eq:vminN}
\end{equation}
which reduces to the elastic case for $\delta=0$. Requiring
$\vmin(E_R)\leq\vmax$, where $\vmax$ is the largest DM speed in the
laboratory frame, gives~\citep{Herrera:2023nlp}
\begin{equation}
\delta \leq \vmax\sqrt{2m_T E_R}
-\frac{m_T E_R}{\mu_{\chi T}}
\leq \frac{1}{2}\mu_{\chi T}\vmax^2 ,
\label{eq:deltaER}
\end{equation}
which directly links the accessible splitting to the fastest incoming
particles. The first inequality applies at a given recoil energy. The second is the absolute ceiling: as a function of $E_R$, the middle expression is largest at $E_R=\mu_{\chi T}^2\vmax^2/(2m_T)$, where it equals $\tfrac{1}{2}\mu_{\chi T}\vmax^2$, so no recoil energy makes a larger splitting accessible.\footnote{A similar bound applies to DM--electron scattering. There, however, the energy deposited, $\Delta E$, is set by the electronic transition rather than by the momentum transfer, so the two are independent and the minimum speed cannot be lowered below $\sqrt{2\Delta E/m_\chi}$ by any choice of momentum transfer. The low-mass reach is then bounded by $m_\chi\gtrsim2E_{\rm thr}/\vmax^2$, with $E_{\rm thr}$ the detector threshold; see Appendix~\ref{app:electron}.}

In all of these cases, both the population of the high-velocity tail and its laboratory-frame endpoint $\vmax$ are critical inputs. The SHM imposes a sharp cutoff at the Galactic escape speed $\vesc$. The inference of $\vesc$ itself depends on the modeling of the stellar high-speed tail and its kinematic substructure~\citep{Necib:2021yhq,Necib:2021vxr,Roche_2024}. Simulation-based DM distributions can depart from the sharply truncated Maxwellian form and develop additional high-speed structure through the recent infall of the LMC.

The dependence of direct-detection results on the assumed velocity distribution has been studied extensively, most commonly for the standard spin-independent interaction. Halo-independent methods were developed to compare experiments without assuming a functional form for $f(\mathbf{v})$~\citep{Fox:2010bz,Fox:2010bu,Gondolo:2012rs,Herrero-Garcia:2012arz,DelNobile:2013cta,Frandsen:2013cna,Feldstein:2014gza,Bozorgnia:2014gsa,Blennow:2015oea,Anderson:2015xaa,Gelmini:2016pei}, and were subsequently extended to generalized interactions and the NREFT operator basis~\citep{DelNobile:2013cva,Kahlhoefer:2016eds,Catena:2018ywo,Kang:2022zqv,Choi:2024rmq}, as well as to inelastic kinematics~\citep{Kang:2023gef}. Complementary approaches have parametrized departures from the Maxwellian form non-parametrically~\citep{Fowlie:2018svr} or through information divergences~\citep{Herrera:2024zrk,Herrera:2026lqi}, finding that the resulting uncertainty can become particularly large for momentum- and velocity-dependent interactions near kinematic thresholds. The consequences of kinematic substructure, the LMC, and halo-to-halo variation have likewise been explored for specific targets and interactions~\citep{OHare:2018trr,Herrera:2021puj,Maity:2022enp,Reynoso-Cordova:2024xqz,Bozorgnia:2025lsl,Kakharov:2025myy,Folsom:2025lly,Lilie:2025wkr}; for reviews see Refs.~\citep{Green:2011bv,Green:2017odb}.

The same sensitivity that affects exclusion limits also enters the interpretation of a potential signal. A particularly timely example is provided by LUX-ZEPLIN, whose extended nuclear-recoil search reported a single event at $248 \pm 23~({\rm stat}) \pm 23~({\rm sys})~$keV above a background expectation of $0.0106$ events~\citep{LZ:2026ext}. Because such a high-energy recoil probes unusually large incoming DM speeds, an inelastic interpretation lies directly in the tail-dominated regime of Eq.~\eqref{eq:deltaER}. The mass splitting inferred from the event can therefore depend as strongly on the assumed high-velocity distribution as on the underlying DM interaction itself.

The remainder of this paper brings these astrophysical and particle physics questions together. In Sec.~\ref{sec:vdf}, we apply the reconstruction framework developed and calibrated in Ref.~\citep{zhang2026setnightfirebuilding} to \textit{Gaia} DR3. Gaia--Sausage--Enceladus (GSE) is a major merger in the MW's assembly history that produced the dominant accreted stellar population in the local halo~\citep{2018MNRAS.478..611B,2018Natur.563...85H, Myeong_2018, 2019MNRAS.486..378L, Zhu_2024}. We use stars associated with GSE to infer the kinematics of the corresponding DM population and thereby reconstruct the local MW velocity distribution. We then incorporate the closest LMC analog in FIRE-2 and compare the resulting high-velocity tail with the two commonly used benchmarks, the SHM and Au-25, in Sec.~\ref{subsec:LMC}. In Sec.~\ref{sec:eft}, we introduce the non-relativistic effective theory used to describe the DM--nucleon interactions. In Sec.~\ref{sec:impact}, we propagate the different halo models through the full operator basis for both elastic and inelastic scattering, and use the high-energy event reported in the LUX-ZEPLIN extended search as a concrete example of how the assumed velocity distribution affects the interpretation of a direct-detection signal. We conclude in Sec.~\ref{sec:conclusions}, and additionally consider DM--electron scattering in the App.~\ref{app:electron}.

\section{Local DM Velocity Reconstruction}
\label{sec:vdf}

In this section, we describe our framework for inferring the local DM velocity distribution using a reconstruction methodology established and validated with cosmological simulations~\citep{zhang2026setnightfirebuilding} and subsequently applied to observational data. In Sec.~\ref{subsec:FIRE}, we first introduce the FIRE-2 simulation suite employed in this work. In Sec.~\ref{subsec:method}, we review the reconstruction methodology developed in our previous study~\citep{zhang2026setnightfirebuilding}. We then apply this framework to two major merger-related components of the MW halo, the GSE and the LMC. In Sec.~\ref{subsec:GSE}, we discuss the GSE merger and infer its contribution to the local DM velocity distribution from observational data. Finally, in Sec.~\ref{subsec:LMC}, we examine the impact of the LMC on the high-velocity tail of the local DM distribution.

\subsection{FIRE simulations} \label{subsec:FIRE}

The FIRE \textit{Latte} suite consists of cosmological zoom-in hydrodynamic simulations that adopt the Feedback In Realistic Environments (FIRE)-2 galaxy-formation model. The simulations are run with \textsc{GIZMO} using the meshless finite-mass (MFM) hydrodynamics method~\citep{2014MNRAS.445..581H, Hopkins_2015, 2016ApJ...827L..23W, Hopkins_2018}. Gas is allowed to cool and heat over temperatures of $\sim10$--$10^{10}~\mathrm{K}$ through a range of physical processes, including primordial and metal-line cooling, Compton and free--free emission, low-temperature molecular and dust cooling, and photoheating from a redshift-dependent ultraviolet background, with local self-shielding included~\citep{2009ApJ...703.1416F}. Star formation occurs only in dense, locally self-gravitating gas, with a typical density threshold of $n_{\rm H}\ge10^3,\mathrm{cm^{-3}}$. Gas satisfying these criteria is converted into star particles on the local free-fall timescale, while the overall star-formation efficiency is regulated by stellar feedback. Each star particle represents a single-age stellar population sampled from a Kroupa initial mass function~\citep{2001MNRAS.322..231K}. Stellar feedback is treated in a time-resolved manner and includes Type II and Type Ia supernovae, OB and AGB stellar winds, photoionization and photoelectric heating, and radiation pressure. Mass, metals, energy, and momentum are coupled conservatively to neighboring gas elements, with multiple metal species tracked together with subgrid turbulent metal diffusion.

The simulated systems are selected to resemble isolated MW-mass halos, with virial masses of $\sim10^{12}M_\odot$ and no halo of comparable mass within five virial radii.\footnote{We define the virial radius $R_{200}$ as the radius enclosing a mean density of $200\rho_{\mathrm{c}}$, where $\rho_{\mathrm{c}}=3H(t)^2/(8\pi G)$, and define the virial mass as the total mass enclosed within $R_{200}$.} The Plummer-equivalent gravitational softenings are $4~\mathrm{pc}$ for star particles and $40~\mathrm{pc}$ for DM particles, while the adaptive gas softening reaches a minimum of $1~\mathrm{pc}$.

\subsection{Reconstruction method} \label{subsec:method}

The DM halo is commonly modeled using the Standard Halo Model (SHM)~\citep{1986PhRvD..33.3495D, 1996APh.....6...87L}, which assumes a spherical, isotropic, and time-independent halo in dynamical equilibrium. Under these assumptions, the Galactic-frame DM velocity distribution is taken to be Maxwell--Boltzmann, such that each velocity component follows a Gaussian distribution. The distribution is truncated at the Galactic escape speed, typically $v_{\rm esc}\sim 500-600~{\rm km/s}$~\citep[see e.g.][]{Smith_2007, Monari_2018, 2021A&A...649A.136K,2022ApJ...926..189N,2024ApJ...972...70R}, and subsequently boosted into the Earth's rest frame. In realistic MW-like halos, however, these assumptions can be violated by merger events. In this section, we review the methodology developed in \citet{zhang2026setnightfirebuilding} for reconstructing the local DM velocity distribution in the presence of such merger-driven perturbations.
 
In \citet{zhang2026setnightfirebuilding}, the local DM population is decomposed into two broad components: a merger component associated with progenitors that contribute substantial fractions of the local star density, and a diffuse component containing the remaining material. The latter includes smoothly accreted DM as well as debris from lower-mass mergers.
 
For the merger component, the local DM velocity distribution can be inferred from the accreted stellar population associated with the same progenitor. Simulations show that the stellar and DM debris trace similar kinematic structures, while the DM component generally exhibits a broader velocity dispersion~\citet{zhang2026setnightfirebuilding}. This difference can be modeled by convolving the stellar velocity distribution with a Gaussian kernel whose width accounts for the additional DM velocity dispersion.
 
The diffuse component, which contains smoothly accreted DM and smaller mergers, is therefore expected to be more dynamically relaxed and closer to the SHM expectation. Nevertheless, \citet{zhang2026setnightfirebuilding} showed that its velocity distribution is more accurately described by a generalized Gaussian, reflecting residual departures from equilibrium induced by interactions with mergers. For each velocity component, the distribution is modeled as
\begin{equation}
\label{eq:gengauss}
f_g(v) =
\frac{\beta}{2\alpha\Gamma\left(1/\beta\right)}
\exp\left[
-\left|\frac{v-\mu}{\alpha}\right|^\beta
\right],
\end{equation}
where $\mu$ is the location parameter, $\alpha$ sets the characteristic width of the distribution, and $\beta$ controls its shape. The Gaussian limit is recovered for $\beta=2$, while deviations from this value allow the model to capture non-Gaussian features in the local velocity distribution. Simulations indicate that the best-fit location parameter is consistent with $\mu=0$ for all three velocity components~\citep{zhang2026setnightfirebuilding}. We therefore fix $\mu=0$ and adopt the simulation-calibrated best-fit values of $\alpha$ and $\beta$ reported in \citet{zhang2026setnightfirebuilding}, which are summarized in Table~\ref{tab:best_fit}.

\begin{table}[t]
\centering
\setlength{\tabcolsep}{14pt}
\begin{tabular}{lcc}
\hline\hline
 & \textbf{$\beta$} & \textbf{$\alpha$/$v_c$}\\
\hline
$r$ & 2.74 $\pm$0.31 & 1.17\\
$\theta$ & 2.37 $\pm$0.20 & 1.05 $\pm$0.07 \\
$\phi$ & 2.35 $\pm$0.22 & 1.08\\
\hline
\end{tabular}
\caption{\label{tab:best_fit}Simulation-calibrated best-fit parameters of the generalized Gaussian model for the diffuse local DM velocity distribution, as obtained in \citet{zhang2026setnightfirebuilding}. The parameters $\alpha$ and $\beta$ characterize the width and shape of the distribution, respectively, for each velocity component, with the location parameter fixed to $\mu=0$.}
\end{table}

The local DM velocity distribution is reconstructed by combining the diffuse and merger components. The diffuse component is modeled using the generalized Gaussian distribution described above, with the simulation-calibrated best-fit parameters listed in Table~\ref{tab:best_fit}. We adopt a local circular velocity of $v_c = 233 \pm 5~{\rm km,s^{-1}}$~\citep{1979IAUS...84..231E, Camarillo_2018, ou2023darkmatterprofilemilky}. For the merger component, we focus on GSE, the most recent major accretion event. Earlier mergers were accreted sufficiently long ago that their debris is expected to be largely phase mixed and is therefore incorporated into the diffuse component. More recent accretion events, most notably the LMC, are still in the early stages of infall~\cite{2013ApJ...764..161K, 2017MNRAS.464.3825P, 2019MNRAS.487.2685E, 2024MNRAS.534.2694S} and are treated separately rather than as part of the local merger component. In the following sections, we first describe our reconstruction of the GSE contribution and then examine the impact of the LMC on the high-velocity tail of the local DM distribution.

\subsection{\textit{Gaia} Sausage Enceladus} \label{subsec:GSE}

The GSE is the remnant of a major merger experienced by the MW approximately 8--10 Gyr ago and is characterized by debris on highly radial orbits~\citep{2018MNRAS.478..611B,2018Natur.563...85H}. Estimates of the merger--to--host mass ratio are typically of order 1{:}4 to 1{:}8~\citep{2018Natur.563...85H, Bignone_2019, 2023MNRAS.526.1209L}. Its stellar debris is thought to constitute roughly half of the nearby stellar halo, with some studies suggesting that it may account for as much as $85\%$ of the local accreted, metal-poor stellar population~\citep{Myeong_2018, 2019MNRAS.486..378L, Zhu_2024}. To determine the stellar kinematics of the GSE, we follow the methodology developed in \citet{Ostdiek_2020,Necib_2020}. In \citet{Ostdiek_2020}, a deep neural network was trained to distinguish accreted from in-situ stars using phase-space information, producing a catalog of candidate accreted stars in Gaia DR2. \citet{Necib_2020} subsequently analyzed stars with full three-dimensional velocity information in the Solar neighborhood, defined by $6.5<r<9.5~{\rm kpc}$ and $|z|<3~{\rm kpc}$, and applied a Gaussian mixture model in Galactocentric velocity space to separate distinct accreted stellar populations. In this work, we apply the same analysis pipeline to Gaia DR3 to infer the GSE stellar kinematics.

\begin{figure}[t]
\centering
\includegraphics[width=0.5\textwidth]{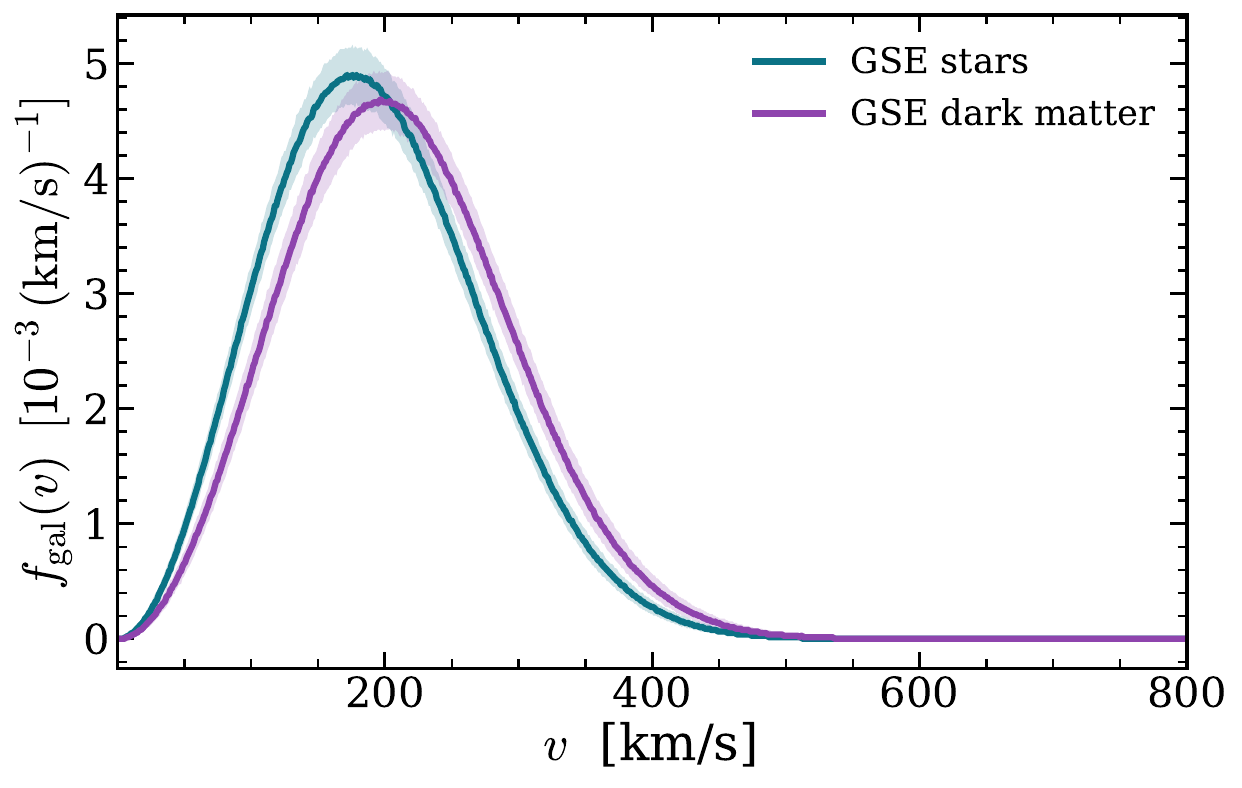}
\caption{Velocity distribution of the GSE component in the Solar neighborhood. The blue curves show the stellar velocity distribution inferred from Gaia DR3 using the Gaussian mixture model described in the text, while the purple curves show the corresponding GSE-associated DM distribution reconstructed by convolving the stellar distribution with the simulation-calibrated Gaussian kernel of \citet{zhang2026setnightfirebuilding}. The solid curves denote the median distributions over the ensemble of realizations, and the shaded bands indicate the corresponding $1\sigma$ containment regions.}
\label{fig:GSE}
\end{figure}

The GSE velocity distribution is modeled with a Gaussian mixture model in Galactocentric spherical velocity coordinates. The two angular velocity components are each described by a single Gaussian centered near zero, while the radial component is modeled as a bimodal Gaussian distribution to capture the strongly radial nature of the GSE debris. We sample the best-fit means and dispersions, together with their corresponding uncertainties, to propagate the observational uncertainties in the inferred stellar kinematics. For each realization of the model parameters, we generate $10^4$ velocity vectors. This procedure is repeated for $8\times10^5$ independent realizations, allowing us to characterize the uncertainty in the resulting velocity distribution. Following \citet{zhang2026setnightfirebuilding}, the velocity distribution of the GSE-associated DM component is then reconstructed from the observed stellar velocity distribution by convolving it with a Gaussian kernel. The width of this kernel is calibrated using simulations to account for the systematically broader velocity distribution of the DM debris relative to its stellar counterpart. The resulting stellar and inferred DM velocity distributions are shown in Fig.~\ref{fig:GSE} as the blue and purple curves, respectively, with the shaded bands indicating the $1\sigma$ containment region across the ensemble of realizations.

The fraction of the local DM density contributed by GSE is estimated as
\begin{equation}
    f_{\rm GSE}
    =
    \gamma'
    \frac{m_{\rm DM}^{\rm peak}}{m_{*}^{\rm peak}}
    \frac{f_{\rm GSE}^{\rm acc}\rho_{*}^{\rm acc}}
         {\rho_{\rm DM}^{\odot}} .
\end{equation}
Here, $\gamma'$ is a correction factor that relates the global DM-to-stellar mass ratio of the progenitor at peak mass, $m_{\rm DM}^{\rm peak}/m_{*}^{\rm peak}$, to the corresponding ratio of DM and stellar debris deposited in the Solar neighborhood. This factor is calibrated using simulations in \citet{zhang2026setnightfirebuilding} and depends primarily on the accretion redshift of the progenitor, with typical values of order $\mathcal{O}(0.1)$.

For the GSE, the inferred accretion lookback time is $9.5^{+0.8}_{-1.4}$ Gyr, corresponding to an accretion redshift of approximately $z_{\rm acc}=1.49^{+0.38}_{-0.45}$~\citep{sante2026simulationbasedinferencemilkyway}. To propagate this uncertainty, we model the accretion lookback time using a split Gaussian distribution \[
f(x;\mu,\sigma_1,\sigma_2)
=
\begin{cases}
A\exp\left(
-\dfrac{(x-\mu)^2}{2\sigma_1^2}
\right),
& \text{if } x<\mu, \\[1ex]
A\exp\left(
-\dfrac{(x-\mu)^2}{2\sigma_2^2}
\right),
& \text{otherwise},
\end{cases}
\] and convert each sampled lookback time to redshift assuming the fiducial cosmology. The progenitor DM-to-stellar mass ratio at peak mass, $m_{\rm DM}^{\rm peak}/m_{*}^{\rm peak}$, is then inferred from the stellar mass--halo mass (SMHM) relation of \citet{Behroozi_2019}. The quantity $f_{\rm GSE}^{\rm acc}$ denotes the fraction of the local accreted stellar population associated with GSE, for which we adopt $f_{\rm GSE}^{\rm acc}=0.85$ following \citet{Zhu_2024}. Finally, $\rho_{*}^{\rm acc}$ is the local mass density of accreted stars~\citep{Everall_2022}, while $\rho_{\rm DM}^{\odot}$ denotes the total local DM density.

\begin{figure}[t]
\centering
\includegraphics[width=0.5\textwidth]{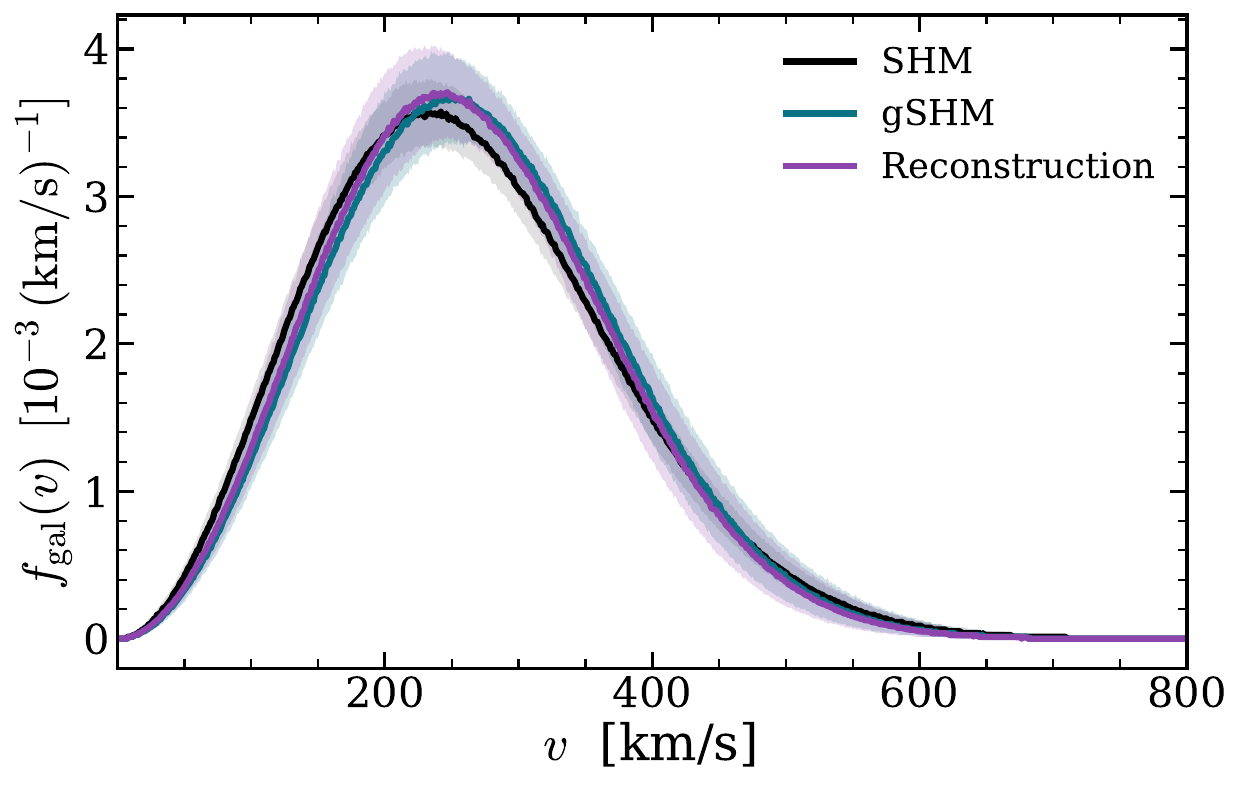}
\caption{Reconstructed local DM speed distribution in the Solar neighborhood. The purple curve shows the full reconstructed distribution, including both the diffuse and GSE-associated components. For comparison, the SHM and gSHM predictions are shown in black and blue, respectively. }
\label{fig:MW}
\end{figure}

The reconstructed local DM velocity distribution is shown in Fig.~\ref{fig:MW} as the purple curve, together with the SHM and generalized SHM (gSHM) predictions shown in black and blue, respectively. The shaded $1\sigma$ bands associated with the SHM reflects the uncertainty in the adopted local circular velocity. Relative to the SHM, the gSHM distribution peaks at a slightly higher speed. The inclusion of the GSE component partially offsets this shift, since the GSE debris is kinematically colder and preferentially contributes at somewhat lower speeds than the diffuse component. As a result, the peak of the reconstructed distribution remains shifted to higher speed relative to the SHM, but lies closer to the SHM prediction than does the gSHM distribution alone, which is consistent with other studies~\cite{Shpigel_2026}.

\subsection{LMC and high velocity tail} \label{subsec:LMC}

\subsubsection{LMC and the Auriga Benchmark}\label{subsubsec:m12bLMC}

In the previous section, we described how the local DM velocity distribution can be reconstructed using stellar tracers of accreted populations. The high-velocity tail of the distribution, however, remains considerably less well constrained. This regime is particularly important for direct-detection experiments, since the fastest DM particles can dominate the scattering rate for low-mass DM and for scenarios involving inelastic scattering with a nonzero kinematic threshold. One contribution that therefore requires explicit treatment is the LMC. 

The LMC is the most massive satellite galaxy of the MW, with an inferred halo mass of order 10--20$\%$ of that of the MW~\citep{2016MNRAS.456L..54P, 2019MNRAS.487.2685E}. It is thought to have entered the MW halo only $\sim1.5-2$ Gyr ago and to have recently undergone its first pericentric passage 50 Myr ago; it is currently located approximately $50~{\rm kpc}$ from the Galactic center~\citep{2013ApJ...764..161K, 2017MNRAS.464.3825P, 2019MNRAS.487.2685E, 2024MNRAS.534.2694S}. Because of its recent infall, almost all of the LMC DM is expected to remain kinematically associated with its center-of-mass motion rather than being phase mixed with the MW halo. Moreover, owing to its large pre-infall halo mass, the LMC DM halo is expected to have extended to radii well above $100~{\rm kpc}$. It is therefore plausible that LMC-associated DM contributes to the local DM population near the Solar neighborhood. Indeed, numerical studies have shown that, although the LMC typically contributes only a small fraction of the total local DM density, often at the sub-percent level, its debris dominates the high-velocity tail of the local speed distribution~\citep{2019ApJ...884...51G, 2022MNRAS.513L..46D, SmithOrlik:2023pbo}.

Most of the recent literature in inelastic DM has relied on results from a particular LMC analog in the Auriga simulation suite~\cite{Bozorgnia:2024kkf, Graham:2024syw, DallaValleGarcia:2025cwf, Graham:2026ivn, Fan:2026kxx, Smirnov:2026aqk, Lee:2026xxh, Asadi:2026iot, He:2026hqz, Palmisano:2026kuj, Lian:2026hpm, Ghosh:2026txe, Bozorgnia:2026emp, Chanda:2026wdl, Freese:2026mfi, Langhoff:2026qgo, cerdeno2026}, Au-25, which predicts LMC-associated particles with heliocentric speeds extending to approximately $950~{\rm km/s}$~\citep{SmithOrlik:2023pbo}. The Auriga simulation suite is another set of cosmological simulations, each of which has a distinct assembly history and therefore not necessarily host an LMC-analog. To enlarge the sample of MW-LMC analogues, the analysis~\cite{SmithOrlik:2023pbo} searches the merger histories of the simulated hosts back to $z=1$. Candidate systems are selected by requiring that (i) the stellar mass of the LMC analogue exceed $5\times 10^8 M_{\odot}$; (ii) the distance to the galactic center at first pericenter is between 40 to 60~kpc. Based on these criteria, 15 MW-LMC pairs are identified. The analysis then focuses on Au-25, whose merger to host ratio is 0.27, closest to the observed upper limit of the LMC mass~\cite{Moster_2012, Pe_arrubia_2015, Laporte_2017, Erkal_2019}. The orbit of the Au-25 analogue also differs from that inferred for the real LMC. The first pericentric distance of Au-25 analog is $32.9~{\rm kpc}$, smaller than that of the observed LMC with $50~{\rm kpc}$~\citep{2013ApJ...764..161K, 2017MNRAS.464.3825P, 2019MNRAS.487.2685E, 2024MNRAS.534.2694S}, and it takes approximately $133~{\rm Myr}$ after pericenter to reach a Galactocentric distance of $50~{\rm kpc}$, longer than the 50~Myrs inferred for the real system~\citep{2013ApJ...764..161K, 2017MNRAS.464.3825P, 2019MNRAS.487.2685E, 2024MNRAS.534.2694S}. These differences make it difficult to simultaneously match the present-day relative position and velocity of the Sun and LMC using a Solar analogue confined to the Galactic disk. The analysis therefore allows the Solar position to move out of the disk plane in order to reproduce the desired Sun--LMC phase-space geometry. It should be noted that this assumption is generally valid if the DM halo is isotropic. However, as emphasized in the study ~\citet{Folsom:2026dqs}, simulated DM halos are generally not isotropic and can exhibit an ubiquitous co-rotation. The resulting heliocentric velocity distribution can therefore depend on the adopted Solar position. Independent studies using controlled, non-cosmological MW--LMC simulations have likewise found LMC-associated particles with speeds exceeding the Galactic escape velocity, although the fastest particles in these studies typically lie in the range $700-900~{\rm km/s}$, below the $\sim950~{\rm km/s}$ tail at present day found in Au-25\citep{2019ApJ...884...51G, 2022MNRAS.513L..46D}.

\begin{figure}[t]
\centering
\includegraphics[width=0.45\textwidth]{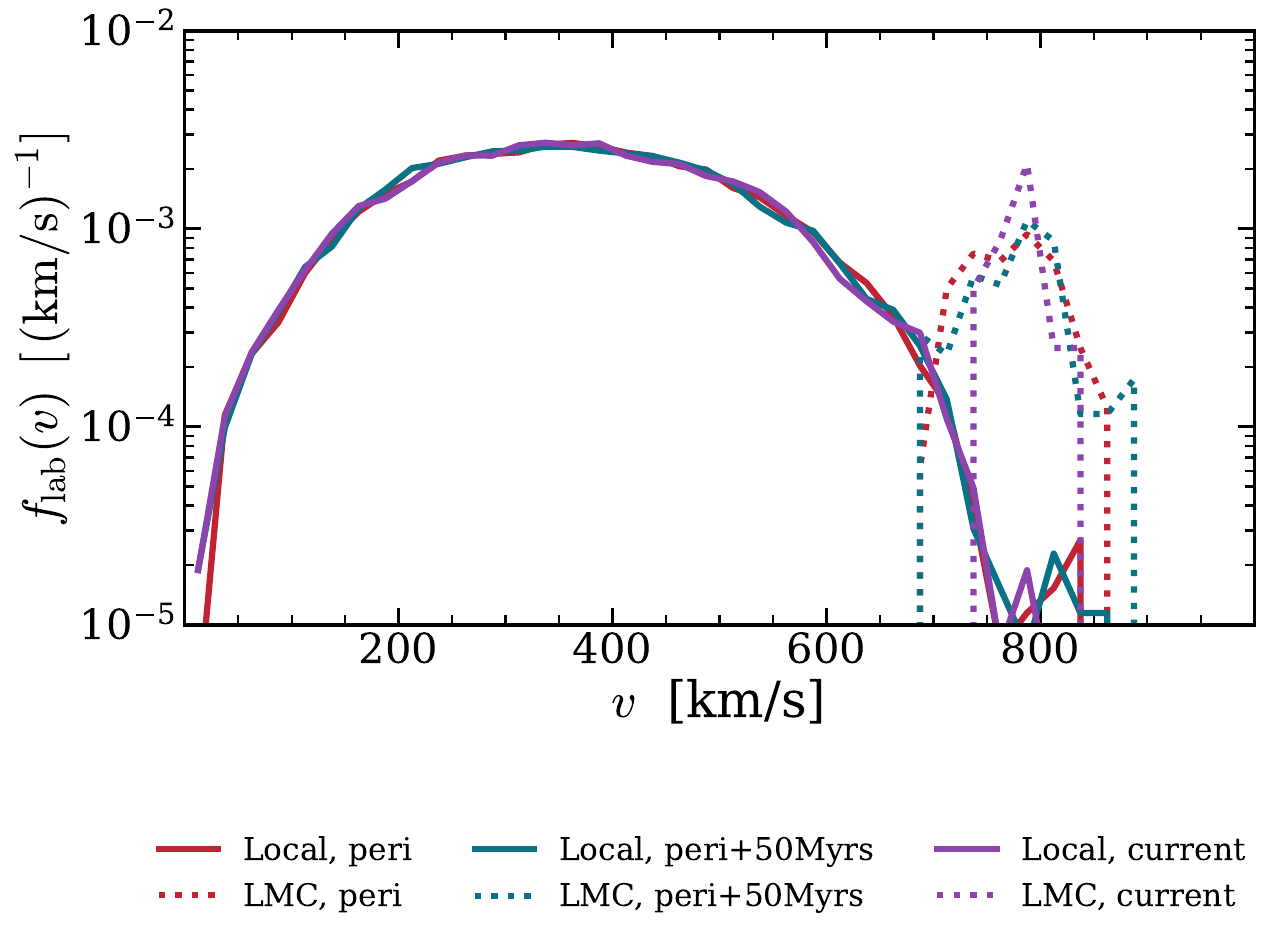}
\caption{MW and LMC contributions to the local DM velocity distribution in the heliocentric frame for the m12b LMC analogue. The distributions are shown for a few selected snapshots: the first pericentric passage at snapshot 384 shown in red, 50Myrs after the first pericenter at snapshot 386 shown in blue, and snapshot 390 shown in purple when the LMC analog reaches a Galactocentric distance of $50~{\rm kpc}$, which we define as the present-day configuration.}
\label{fig:LMC}
\end{figure}

For comparison, in this study we examine an LMC analog in the FIRE-2 simulations. The FIRE LMC analogs were first introduced in \citet{2021MNRAS.504.1379S} with the following selection criteria: (i) the pericenter redshift is less than 0.7; (ii) the subhalo peak mass is greater than $4\times 10^{10} M_{\odot}$; (iii) the stellar mass of the satellite greater than $10^9M_{\odot}$; (iv) the pericenter distance is less than 50kpc. Under these assumptions, 4 LMC analogs were identified in m12b, m12c, m12f, and m12w. However, the pericenter of the m12c and m12w analogs are too close to the galactic center (at 18~kpc and 8~kpc), while the bound mass of the LMC analog in m12f is quite small~\cite{2021MNRAS.504.1379S}. Thus the best LMC analogue in FIRE is the late-time merger in m12b, which is available in the FIRE public release~\citep{Wetzel_2023, wetzel2025secondpublicdatarelease}. It enters the host halo at approximately $z=0.5$, with a merger-to-host mass ratio of 0.12 at current day configuration~\citep{2023MNRAS.523..428B}. The analogue has a pericentric distance of $38~{\rm kpc}$ and takes approximately $150~{\rm Myr}$ to move back out to a Galactocentric distance of $50~{\rm kpc}$. We identify the LMC pericentric passage at snapshot 384 at $z\simeq 0.5$ and follow the system until the analogue moves back out to a Galactocentric distance of $50~{\rm kpc}$ at snapshot 390, which we define as the present-day configuration. To match the present-day relative phase-space geometry, we follow the previous treatment of Au-25 and likewise allow the Solar System to move out of the disk plane. We choose the Solar position and velocity such that, at snapshot 390, the relative phase-space configuration between the Solar System and the merger matches that of the observed MW--LMC system. We then track all DM particles within $2~{\rm kpc}$ of the selected Solar position\footnote{Note that the definition of the Solar neighborhood adopted here differs from that used for Au-25 in Ref.~\cite{SmithOrlik:2023pbo}. To obtain sufficient particle statistics, Ref.~\cite{SmithOrlik:2023pbo} employs a substantially larger sampling volume, which may therefore be less localized to the immediate vicinity of the Sun. Specifically, their Solar neighborhood region is defined as the intersection of a cone originating at the Galactic center, with opening angle $\pi/4$ and axis pointing toward the Solar position, and a spherical shell spanning Galactocentric radii of 6--10 kpc. A more detailed description is given below Fig.~2 of Ref.~\cite{SmithOrlik:2023pbo}.}. As FIRE is also a cosmological simulation, the same caveat regarding halo co-rotation applies to m12b and should therefore be kept in mind when interpreting the heliocentric velocity distribution. Fig.~\ref{fig:LMC} shows the MW and LMC contributions to the local DM velocity distribution in the heliocentric frame in m12b. Due to the interaction with the LMC, the MW DM distribution is shifted toward higher velocities, with the highest-velocity particles reaching approximately 850--$900~{\rm km/s}$ at each snapshot, consistent with other simulation studies. We therefore adopt the LMC velocity distribution at snapshot 390 as a benchmark distribution to combine with our reconstructed local DM velocity distribution. At this snapshot, the LMC contribution contains 32 particles with heliocentric speeds spanning 738.2--846.4$~{\rm kms}$, corresponding to a local DM fraction of $f_{\rm LMC}=0.3\%$.

\subsubsection{Combining the components} \label{subsec:combine}

In the subsequent study, we compare our reconstructed velocity distribution with two benchmark distributions in forms of halo integral: the SHM with an escape velocity cut off at 544~km/s and the Auriga-25 halo with its own LMC analog~\citep{SmithOrlik:2023pbo}. We compare the four descriptions through the halo integral
\begin{equation}
\eta_0(\vmin)=\int_{|\mathbf{v}|\geq\vmin}\frac{f(\mathbf{v})}{v}\,d^3v
=\int_{\vmin}^{\infty}\frac{f_{\rm lab}(v)}{v}\,dv ,
\label{eq:eta0}
\end{equation}
where $f(\mathbf{v})$ is the laboratory-frame velocity distribution,
normalized to unity, $f_{\rm lab}(v)=v^2\!\int d\Omega\,f(\mathbf{v})$ its
angle-averaged speed distribution, and $\vmin$ the minimum speed required
for a given recoil, Eq.~\eqref{eq:vminN}~\citep{DelNobile:2015rmp}. The halo
integral sets the scattering rate, as discussed in Sec.~\ref{sec:eft}, and
it determines the speed distribution, $f_{\rm lab}(v)=-v\,d\eta_0/dv$. For the latter, we use the laboratory-frame halo integral computed in Ref.~\citep{SmithOrlik:2023pbo}, as described below. For our reconstruction, we also compare two different prescriptions: the one described in Sec.~\ref{subsec:method} on its own with no escape truncation, and the reconstruction combined with the \texttt{m12b} LMC debris of Sec.~\ref{subsubsec:m12bLMC}. 

The recent infall of a massive satellite can also accelerate native MW DM through the dynamical response of the host halo, populating velocities above the nominal Galactic escape speed. In simulations of the MW--LMC system, the most extreme high-velocity tail is typically dominated by DM originating from the LMC itself~\cite{Besla:2019xbx, SmithOrlik:2023pbo}. We therefore enforce self-consistency by first truncating the reconstruction (which includes a generalized Gaussian distribution that goes to infinity) in the \emph{laboratory} frame at the maximum speed of the LMC debris so that the high velocity tail is dominated and determined by the LMC, and then adding the debris with its local density fraction. Because the debris is sampled by few particles, we smooth it with a Gaussian kernel of bandwidth $25\kms$, comparable to the inter-particle spacing in the tail, truncated at the fastest particle so that the smoothing does not extend the distribution beyond the data from the simulation. Without smoothing, the gaps between the last few particles produce zeros in $\eta_0(\vmin)$ that fragment the event-interpretation regions of Sec.~\ref{subsec:islands} into disconnected pieces. The extreme tail of this component is represented by a small number of particles, and the conclusions drawn from it below are limited accordingly.

For Auriga-25 we use directly the laboratory-frame halo integral
$\eta_0(\vmin)$ that Ref.~\citep{SmithOrlik:2023pbo} computes from the simulated
particle velocities, time-averaged over the year, for the present-day
snapshot including the Au-25 LMC analog. $\eta_0$ vanishes at $975\kms$, the upper edge of the last
bin in which it is nonzero.
Since this halo integral is computed from
the particle velocities, it retains the anisotropy of the LMC debris, which
arrives from a preferred direction, and requires no assumption about the
Galactic-frame distribution. The SHM
is transformed to the laboratory frame with the Galilean shift
\begin{equation}
f_{\rm lab}(\mathbf{v}) = f_{\rm gal}(\mathbf{v}+\mathbf{v}_E) ,
\label{eq:boost}
\end{equation}
where $\mathbf{v}_E$ is the velocity of the Earth in the Galactic rest
frame, of magnitude $v_E=|\mathbf{v}_E|=250.2\kms$ when averaged over the
year~\citep{1996APh.....6...87L,Savage:2008er}. Performing the angular
integration numerically reproduces the closed-form SHM halo integrals of
Refs.~\citep{1996APh.....6...87L,Savage:2008er} to better than one part in
$10^{5}$. The reconstruction and the LMC debris are supplied in the
heliocentric frame and require no such transformation. 

\subsubsection{The resulting distributions} \label{subsec:money}

\begin{figure}[t]
\includegraphics[width=0.45\textwidth]{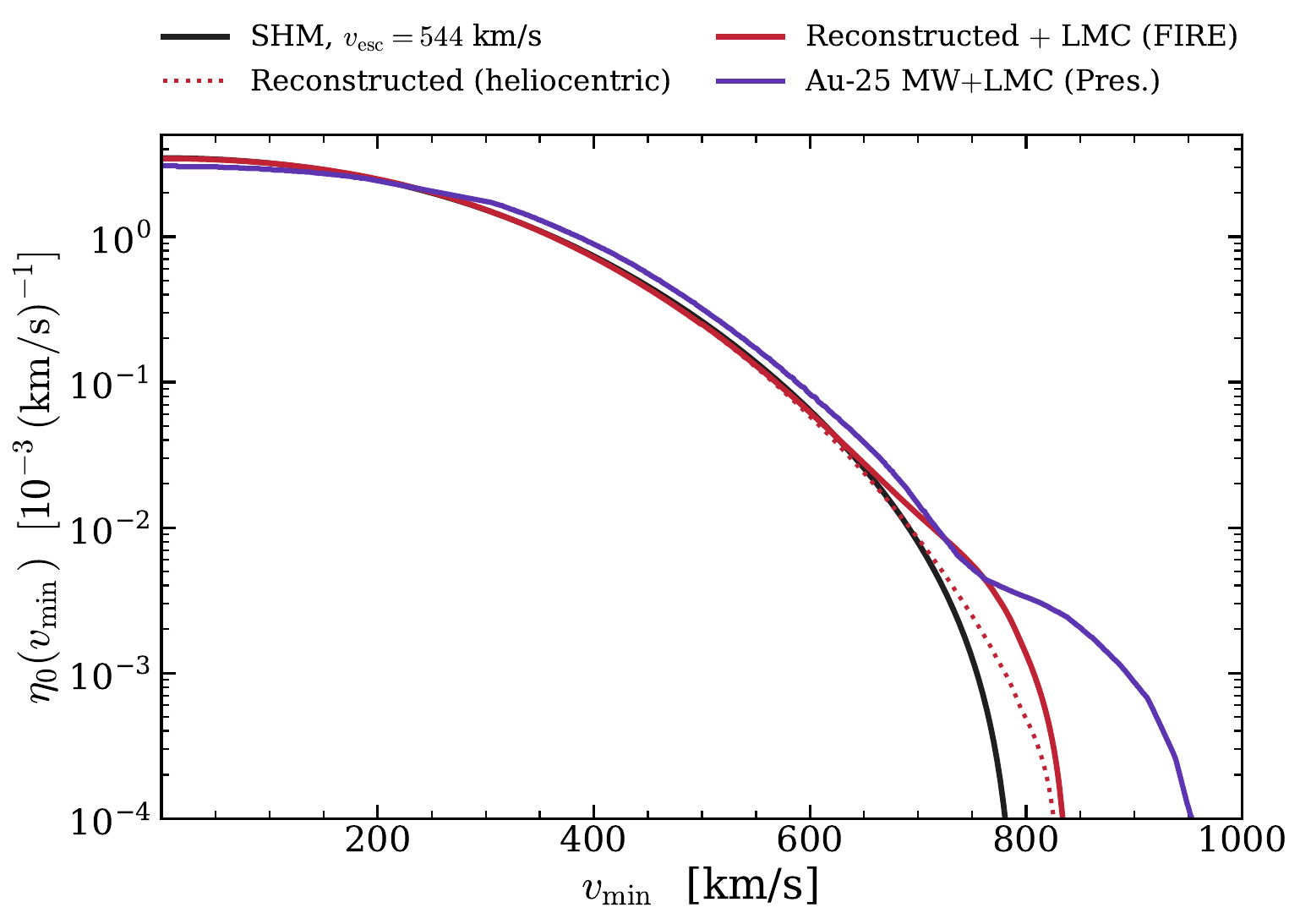}
\caption{Halo integrals for the
simulation-based descriptions, with the SHM as the central benchmark. Note that on a logarithmic axis a tail extending to $f_{\rm lab} \sim 10^{-5}$ or $\eta_{0} \sim 10^{-7}$ appears as prominent as the bulk: the Auriga-25 tail beyond $950\kms$ carries $10^{-4}$ of its normalization, and $99.99\%$ of that distribution lies below $951\kms$ (Table~\ref{tab:endpoints}). The SHM and the FIRE-2-based distributions agree to a few percent below $500\kms$, while Auriga-25 is shifted to higher speeds by $10$--$25\%$ across the bulk; all separate by orders of magnitude above $700\kms$. }
\label{fig:money}
\end{figure}

Figure~\ref{fig:money} summarizes our benchmark distributions that provide the central empirical comparison of this work, and two regimes are apparent. 

Below $600\kms$, the SHM and the FIRE-2-based halo integrals agree closely, to within $1\%$ at $300$ and $5\%$ at $500\kms$. Auriga-25, whose laboratory-frame distribution is shifted to higher speeds, differs from them by $14\%$ at $300\kms$ and $22\%$ at $500\kms$. Observables that integrate over the bulk of the distribution are therefore insensitive to the choice between the SHM and the FIRE-2-based descriptions at the level of a few percent, but can differ by tens of percent for Auriga-25.

The predictions increasingly diverge at high speeds, differing by $29\%$ at $600\kms$ and $84\%$ at $700\kms$. Beyond the lowest endpoint, the ratio becomes formally infinite because some distributions vanish while others remain nonzero. Table~\ref{tab:endpoints} lists the maximum laboratory-frame speed and the corresponding upper limit on the inelastic mass splitting.
\begin{equation}
\delta_{\max} = \tfrac{1}{2}\,\mu_{\chi T}\,\vmax^2 ,
\label{eq:dmax}
\end{equation}
for a $1\,$TeV particle on Xenon. Although LMC debris contributes only $0.30\%$ of the local DM density, it raises the maximum speed above the SHM value. Even this small population can therefore affect the experimental reach more than percent-level changes in the bulk of the distribution.

\subsubsection{Posterior spread of the endpoint} \label{subsec:ensemble}

The reconstruction is delivered as an ensemble of $1000$ posterior
realizations, and every curve shown so far is their bin-wise median. The
median is a convenient summary but is not itself a member of the ensemble,
and in the tail it is biased low by construction: a bin is non-zero in the
median only where at least half of the realizations have non zero bins, so the
median curve ends at $646\kms$ in the Galactic frame while the median
realization reaches $744\kms$ and the fastest $798\kms$. Carried into the
laboratory frame, the median curve stops at $835\kms$, whereas the ensemble
spans $773$--$912\kms$ between the $16$th and $84$th percentiles. One should be cautious in interpreting the behavior of the highest-velocity tail, since it is determined merely by extrapolating the best-fit functional form obtained from the bulk of the velocity distribution. The large spread therefore should not be interpreted as a physical effect, but rather as reflecting the substantial uncertainty in where the distribution ultimately truncates, and should be treated with caution. The m12b LMC result, on the other hand, is broadly consistent with \citet{2019ApJ...884...51G, 2022MNRAS.513L..46D}, both of which find significantly lower LMC-induced velocities than those obtained for Au-25.

We therefore propagate the realizations individually through the same
machinery wherever the spread matters, and show it as a band in
Fig.~\ref{fig:ratio}. Adding the LMC debris removes this ambiguity on the endpoint of the high velocity tail: once the velocity distribution is truncated at the debris endpoint, every realization terminates at $846.4\kms$ by construction, and the endpoint becomes a property of the simulated satellite rather than of the posterior.

\begin{table}[b]
\caption{Laboratory-frame endpoint of each simulation-based description and
the implied ceiling on the inelastic mass splitting, Eq.~\eqref{eq:dmax}, for
$m_\chi = 1\,$TeV on xenon. $\vmax$ is the formal edge of the support, which
is what the rate calculation uses throughout; $v_{99.99}$ is the speed below
which all but $10^{-4}$ of the halo lies, and is the more meaningful measure
of reach where the last bins are sparsely populated. For Auriga-25, whose halo integral is taken directly from Ref.~\citep{SmithOrlik:2023pbo}, $v_{99.99} = 951\kms$ is consistent with the endpoint near $950\kms$ of the laboratory-frame halo integral in Ref.~\citep{SmithOrlik:2023pbo}, and the formal edge, the upper end of the last $25\kms$ bin in which $\eta_0$ is nonzero, lies at $975\kms$. The reconstruction appears twice: the second row is the
bin-wise median curve used in the figures, the third the $16$--$84\%$ range
over its posterior realizations, which quantifies how well the endpoint is
determined (Sec.~\ref{subsec:ensemble}).}
\label{tab:endpoints}
\begin{ruledtabular}
\begin{tabular}{lccc}
description & $\vmax$ & $v_{99.99}$ & $\delta_{\max}$ \\
& [km/s] & [km/s] & [keV] \\
\colrule
SHM, $\vesc = 544\kms$ & 793 & 779 & 381 \\
Reconstruction, no escape cut & 835 & 824 & 423 \\
\quad posterior ensemble, p16--p84 & 773--912 & --- & 362--504 \\
Reconstruction $+$ LMC (m12b FIRE) & 846 & 833 & 434 \\
Auriga-25 $+$ LMC~\citep{SmithOrlik:2023pbo} & 975 & 951 & 575 \\
\end{tabular}
\end{ruledtabular}
\end{table}

\section{Direct detection in the effective theory of DM--nucleon
interactions}
\label{sec:eft}

We now introduce the non-relativistic effective theory of DM--nucleon interactions used to translate the velocity distributions discussed above into direct detection predictions. We summarize the operator basis, scattering kinematics, and velocity integrals that determine how each interaction responds to changes in the high-velocity tail, providing the framework for the experimental comparisons in Sec.~\ref{sec:impact}.
\subsection{The operator basis}

DM particles in the solar neighborhood move with $v\sim\mathcal{O}(100)\, \rm km/s \ll c$, and the momentum transferred to a xenon nucleus, $q=\sqrt{2m_TE_R}\simeq 50$--$300\,$MeV, resolves nuclear structure while remaining below the nucleon mass, so that the scattering is described by the non-relativistic effective theory of Refs.~\citep{Fan:2010gt,Fitzpatrick:2012ix,Anand:2013yka}, valid for mediators heavier than $q$. At the nucleon level the Hamiltonian reads 
\begin{equation}
\hat{\mathcal{H}}_N = \sum_{\tau=0,1}\;\sum_{j}
c_j^{\tau}\,\hat{\mathcal{O}}_j\,t^{\tau} ,
\label{eq:hamN}
\end{equation}
with $t^0 = \mathbf{1}$ and $t^1 = \tau_3$ the isoscalar and isovector nucleon
operators, and the $\hat{\mathcal{O}}_j$ built from the four Galilean
invariants
\begin{equation}
i\frac{\mathbf{q}}{m_N}, \qquad
\mathbf{v}^{\perp} = \mathbf{v} + \frac{\mathbf{q}}{2\mu_{\chi T}}, \qquad
\mathbf{S}_\chi, \qquad \mathbf{S}_N .
\label{eq:invariants}
\end{equation}
Here $\mathbf{q}$ is the momentum transfer and $\mathbf{v}^{\perp}$ the
component of the relative velocity orthogonal to it, which satisfies
$(\mathbf{v}^{\perp})^2 = v^2 - \vmin^2$; thanks to this identity the whole
velocity dependence of the rate can be written in terms of two integrals
over the distribution. For spin-$1/2$ DM the basis contains
fourteen operators, $\Op{1}$ and $\Op{3}$--$\Op{15}$, since $\Op{2}$ is
not generated at leading order.

Evaluating Eq.~\eqref{eq:hamN} between nuclear states and squaring the scattering amplitude gives six nuclear responses, $M$, $\Sigma'$, $\Sigma''$, $\Delta$, $\Phi''$, and $\tilde\Phi'$, together with the interference responses $M$--$\Phi''$ and $\Sigma'$--$\Delta$~\citep{Anand:2013yka}. We consider one EFT operator at a time, neglecting interference between distinct operators, which can modify direct detection signals when multiple interactions contribute \cite{Catena:2015uua,Brenner:2022qku}. Interference between nuclear responses and between isoscalar and isovector couplings is retained. For a given operator $\hat{\mathcal{O}}_j$, we suppress the index $j$ and write $c^\tau\equiv c_j^\tau$. The differential recoil rate then takes the form
\begin{equation}
\frac{dR}{dE_R} = \frac{\rho_\chi}{m_\chi}\frac{m_T}{2\pi}
\sum_{\tau,\tau'=0,1} c^{\tau} c^{\tau'}
\Big[ S_0^{\tau\tau'}\,\eta_0 +
S_1^{\tau\tau'}\,\big(\eta_1 - \vmin^2\eta_0\big)\Big] ,
\label{eq:rateN}
\end{equation}
where $E_R$ is the nuclear recoil energy, $\rho_\chi$ is the local DM mass density, and $m_\chi$ and $m_T$ are the DM and target-nucleus masses, respectively. The indices $\tau,\tau'=0,1$ label the isoscalar and isovector components. The functions $S_0^{\tau\tau'}$ and $S_1^{\tau\tau'}$ collect the nuclear responses and momentum dependence for the chosen operator, multiplying the contributions independent of and proportional to $(v^\perp)^2$, respectively. Their dependence on the momentum transfer $q=\sqrt{2m_T E_R}$ is implicit. The astrophysical velocity dependence enters through two halo integrals,
\begin{equation}
\begin{aligned}
\eta_0(v_{\min}) &= \int_{|\mathbf{v}|\geq v_{\min}} \frac{f(\mathbf{v})}{v}\,d^3v , \\
\eta_1(v_{\min}) &= \int_{|\mathbf{v}|\geq v_{\min}} v\,f(\mathbf{v})\,d^3v .
\end{aligned}
\label{eq:integrals}
\end{equation}
where $f(\mathbf{v})$ is the laboratory-frame DM velocity distribution, normalized to $\int d^3v\,f(\mathbf{v})=1$, and $v=|\mathbf{v}|$. The integration includes only particles whose speeds exceed the minimum required to produce a recoil, $\vmin(E_R)$. Thus, $\eta_0$ and $\eta_1$ weight the velocity distribution by the inverse speed and the speed, respectively. The first of these is the halo integral of Eq.~\eqref{eq:eta0}. The second is fixed by the first: inserting $f_{\rm lab}(v)=-v\,d\eta_0/dv$ and integrating by parts~\citep{DelNobile:2015rmp},
\begin{equation}
\begin{split}
\eta_1(\vmin)
&=-\int_{\vmin}^{\infty}v^2\,\frac{d\eta_0}{dv}\,dv\\
&=\vmin^2\,\eta_0(\vmin)+2\int_{\vmin}^{\infty}v\,\eta_0(v)\,dv .
\end{split}
\label{eq:eta1eta0}
\end{equation} 
We use this relation for Auriga-25~\citep{SmithOrlik:2023pbo}.

\subsection{Kinematics}

The minimum speed of Eq.~\eqref{eq:vminN} is, for $\delta>0$, not
monotonic. It has a minimum at
\begin{equation}
E_R^{\star} = \frac{\mu_{\chi T}\,\delta}{m_T} ,
\qquad
\vmin^{\star} = \sqrt{\frac{2\delta}{\mu_{\chi T}}} ,
\label{eq:estar}
\end{equation}
and rises on both sides of it. Scattering requires $\vmin^\star<\vmax$. At a specified recoil energy, the requirement $\vmin(E_R)\leq\vmax$ gives the more restrictive bound of Eq.~\eqref{eq:deltaER}.
These relations have three consequences. First, the absolute ceiling depends only on the DM and target masses and the maximum incoming speed, but the experimental recoil-energy window can impose a stricter bound. The sensitivity below that bound also depends on the event rate, exposure, and detector efficiency. Second, the allowed recoil energies lie in a finite interval containing $E_R^\star$, which narrows as $\delta$ approaches the ceiling. Requiring this interval to contain an observed recoil therefore constrains the DM mass and splitting. Finally, near the ceiling, only particles close to $\vmax$ can scatter, making the predicted rate particularly sensitive to the extreme high-velocity tail, where the velocity distribution is most uncertain.

\subsection{Velocity and momentum dependence}

\begin{table}[t]
\caption{NREFT operators grouped by the explicit powers of $q/m_N$ and
$v^{\perp}$ they carry, and the velocity integrals of
Eq.~\eqref{eq:integrals} on which each depends. The basis is common to both
scattering channels. Tail sensitivity increases down the table.}
\label{tab:ops}
\begin{ruledtabular}
\begin{tabular}{llcc}
Class & Operators & $(q^{n},v^{m})$ & Integrals \\
\colrule
Unsuppressed & $\Op{1},\Op{4}$ & $(0,0)$ & $\eta_{0}$ \\
Momentum supp.\ & $\Op{9},\Op{10},\Op{11}$ & $(1,0)$ & $\eta_{0}$ \\
& $\Op{6}$ & $(2,0)$ & $\eta_{0}$ \\
Velocity supp.\ & $\Op{7},\Op{8},\Op{12}$ & $(0,1)$ & $\eta_{0},\ \eta_{1}$ \\
Doubly supp.\ & $\Op{3},\Op{5},\Op{13},\Op{14}$ & $(1,1)$
& $\eta_{0},\ \eta_{1}$ \\
& $\Op{15}$ & $(2,1)$ & $\eta_{0},\ \eta_{1}$ \\
\end{tabular}
\end{ruledtabular}
\end{table}

Table~\ref{tab:ops} groups the operators by the explicit powers of $q/m_N$
and $v^{\perp}$ they carry, and two mechanisms govern how sensitive each
group is to the velocity distribution. Operators linear in
$\mathbf{v}^{\perp}$ inherit the $\eta_1$ term of Eq.~\eqref{eq:rateN},
whose weighting by $v f(v)$ favors the fast population, in contrast to the
$1/v$ weighting of $\eta_0$. Operators carrying explicit powers of $q$ are
suppressed at low momentum transfer, which pushes their rate toward larger
recoil energies and, through Eq.~\eqref{eq:vminN}, toward larger $\vmin$.
The doubly suppressed operators combine the two. This grouping fixes the
order in which the curves of Sec.~\ref{subsec:ratios} separate from the
SHM, and within a group the halo dependence is the same for every operator,
because the ratio of two rates at fixed operator depends only on $(n,m)$
and on which integrals appear.

\section{Impact on limits and reconstructed parameters}
\label{sec:impact}

LZ has recently reported an extended nuclear-recoil search over
$5.4$--$269.9\,$keV with an exposure of $2.84\,$t\,yr, in which a single
event at $248\,$keV was observed against an expected background of
$0.0106$ events in the surrounding window~\citep{LZ:2026ext}. We use this
search as our test case throughout the section. Our recoil spectra reproduce those of LZ to within about ten percent, but our limits differ from the published ones by up to a factor of about three, in either direction depending on $\delta$. The difference is likely to arise from the likelihood modeling and the treatment of the observed event. LZ performs a profile likelihood ratio test in $\{S1_c,\log_{10}S2_c\}$ space, with a fitted background model and nuisance parameters for the detector response. This likelihood is not public, and we replace it by the simpler counting construction in reconstructed recoil energy described below.

\subsection{From rates to limits} \label{subsec:method_limits}

All results in this section are obtained with the same pipeline. The
differential rate of Eq.~\eqref{eq:rateN} is evaluated with
\textsc{WimPyDD}~\citep{Jeong:2021bpl}, which implements the
Galilean-invariant effective theory for arbitrary DM spin and
inelastic kinematics, with the velocity distribution entering only through
the tabulated integrals $\eta_0$ and $\eta_1$. The nuclear response
functions are those of \citet{Anand:2013yka}, evaluated for the natural
isotopic composition of xenon, and the local density is fixed to
$\rho_\chi = 0.3\,$GeV/cm$^3$ in every case, so that differences between
halo descriptions reflect the velocity distribution alone.

The expected number of events in an energy interval $\mathcal{R}$ follows
from folding the rate with the detector response,
\begin{equation}
N(\mathcal{R}) = \mathcal{E}\!\int_{\mathcal{R}}\! dE_R\;
\epsilon(E_R)\,\frac{dR}{dE_R} ,
\label{eq:counts}
\end{equation}
where $\mathcal{E}$ is the exposure and $\epsilon(E_R)$ the nuclear-recoil
efficiency of the extended search, digitized from the supplementary
material of \citet{LZ:2026ext}. Wherever the measured energy of the event
enters, the spectrum is additionally smeared with the Gaussian resolution
quoted for that recoil, a statistical width scaling as $\sqrt{E_R}$ added
in quadrature to a systematic one.

With a single event in the region of interest we do not attempt a spectral
fit for the exclusion limits. We instead take the most constraining of
three Poisson statements, each built with the unified Feldman--Cousins
construction~\citep{Feldman:1997qc}: one count is observed in a
resolution-width window around the event, which bounds the expectation
there at $4.36$ events at $90\%$ CL, and no counts are observed in the
sidebands below and above that window, which bounds each of them at
$2.44$. The upper limit on the coupling at fixed mass and splitting is
then
\begin{equation}
\big(c_i^{\tau}\big)^2_{90\%} = \min\!\left[
\frac{4.36}{N_{\rm win}},\;
\frac{2.44}{N_{\rm low}},\;
\frac{2.44}{N_{\rm high}} \right] .
\label{eq:limit}
\end{equation}
This construction is conservative, since it discards the information
carried by the relative normalization of the three regions that a joint
likelihood would retain, and it leaves a visible imprint on the limits. As
a piecewise minimum of three smooth functions, the limit is continuous but
its slope is not, and it changes slope wherever a different region becomes
the binding one. The kinks in Fig.~\ref{fig:applimits} are these
crossovers, which for $\Op{1}$ and the SHM occur between $320$ and
$400\,$keV as the binding region passes from the low sideband to the event
window, to the high sideband and back. They appear twice in the ratio
panels, because numerator and denominator switch at different splittings.
The kinks themselves are a feature of the limit-setting procedure, although their position depends on the velocity distribution through the shape of the recoil spectrum.

\subsection{Limits across the operator basis}

\begin{figure}[t]
\includegraphics[width=\columnwidth]{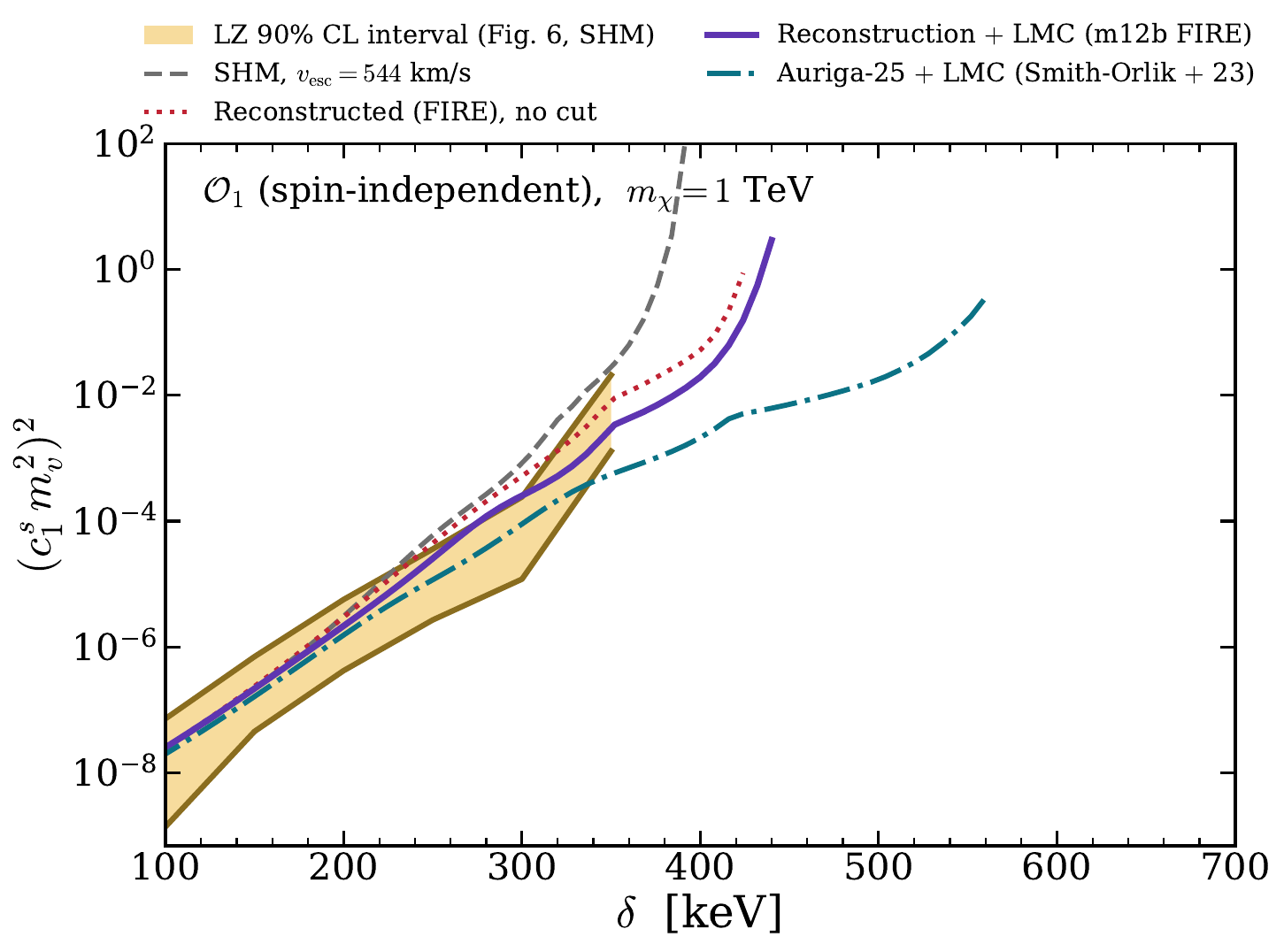}
\caption{$90\%$ CL interval on the isoscalar $\Op{1}$ coupling versus mass
splitting at $m_\chi = 1\,$TeV for each halo description, compared with the
published LZ interval~\citep{LZ:2026ext}. The curves agree below
$\delta \simeq 200\,$keV and separate by orders of magnitude above it,
terminating at the respective endpoints of Table~\ref{tab:endpoints}.}
\label{fig:o1limits}
\end{figure}

\begin{figure*}[p]
\includegraphics[width=\textwidth]{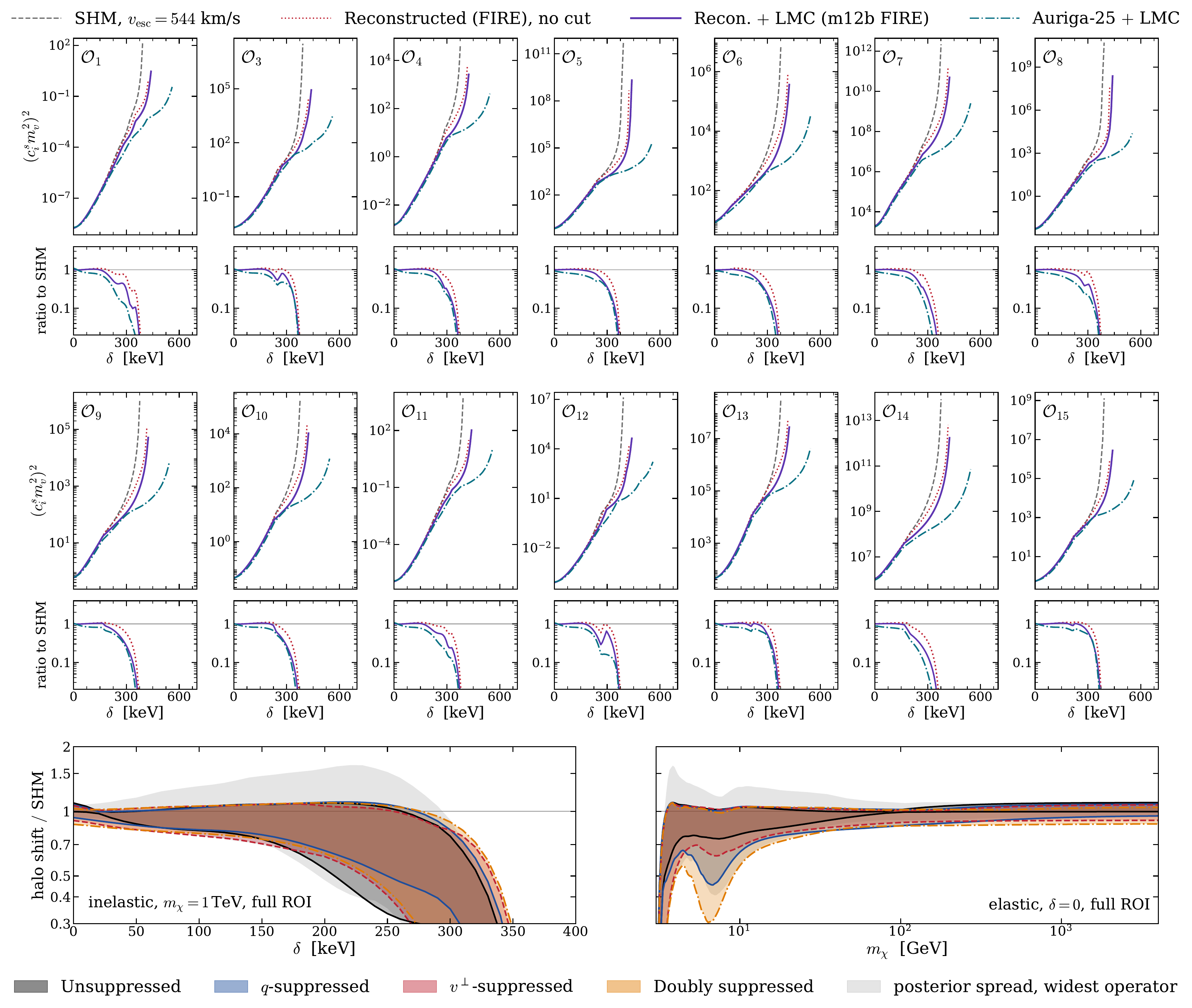}
\caption{Limits and halo dependence for the complete operator basis.
Upper panel of each pair: $90\%$ CL upper limit on the isoscalar coupling
versus mass splitting at $m_\chi = 1\,$TeV, for each halo description,
from the event window and its two sidebands. Lower panel of each pair: the
same limits divided by the SHM one; each curve terminates at the kinematic
ceiling of its own halo, and the ratio tends to zero at the SHM ceiling of
$381\,$keV. Bottom row: the ratio to the SHM over the full region of
interest, compressed into bands by operator class (colours as in
Table~\ref{tab:ops}), inelastic at $1\,$TeV versus $\delta$ (left) and
elastic versus $m_\chi$ (right). Each band spans the full range of the
ratio over the three simulation-based descriptions; the grey band is the
widest $16$--$84\%$ posterior range among the fourteen operators. The
vertical axes of the bottom row are clipped at $0.3$.}
\label{fig:applimits}
\label{fig:ratio}
\end{figure*}

Figure~\ref{fig:o1limits} shows the $90\%$ CL interval on the isoscalar $\Op{1}$ coupling as a function of the mass splitting, computed for the four velocity distributions introduced in Sec.~\ref{sec:vdf}: the SHM, the FIRE-2 reconstruction with and without the LMC debris, and the Auriga-25 halo. Up to $\delta\simeq200\,$keV the four curves coincide, since at these splittings the rate is sourced by the bulk of the distribution, where the descriptions agree. At larger splittings the curves separate, and each terminates at the ceiling set by the endpoint of its distribution. The FIRE-2-based distributions extend the reach of the SHM by some $50\,$keV, and Auriga-25 by close to $200\,$keV. Figure~\ref{fig:applimits} repeats the comparison for the full operator basis.

\subsection{Halo dependence across the operator basis} \label{subsec:ratios}

Fig.~\ref{fig:ratio} shows the limits and halo dependence for all the EFT operators. Each line corresponds to a different velocity distribution prescription. The upper panel of each pair shows the $90\%$ CL upper limit on the isoscalar coupling versus mass splitting at $m_\chi = 1\,$TeV, for each halo description, from the event window and its two sidebands. The lower panel of each pair shows the same limits divided by the SHM limit as the fiducial benchmark; each curve terminates at the kinematic ceiling of its own halo, and the ratio tends to zero at the SHM ceiling of $381\,$keV. A convenient way to read Fig.~\ref{fig:ratio} is through the ratio of the
limit obtained with a given halo description to the limit obtained with the
SHM, at fixed operator, mass and splitting, as shown in the bottom row. A ratio of one means that the
velocity distribution has dropped out of the comparison. The grey band in
the bottom row sets the scale against which departures from one should be
judged: it is the spread between posterior realizations of a single
description, and over most of the range it is as wide as the spread between
operators. Near $\delta = 200\,$keV it is wider, extending from about
$0.5$ to $1.6$ while the operators and descriptions together span
$0.5$--$1.1$.

Two regimes are apparent. For elastic scattering and for small splittings
the ratio stays within ten percent of unity for every operator under the FIRE-2-based descriptions: the tail plays no role and the choice of halo is a subleading systematic. Auriga-25, whose laboratory-frame bulk is shifted to higher speeds, departs from the SHM by up to about twenty percent even there. Above
$\delta \simeq 150\,$keV the curves separate and, for most operators, drop
below unity. In the part of the tail that these splittings sample, every
simulation-based description carries more weight than the SHM, so all of
them strengthen the limit, Auriga-25 most and the reconstruction
without the LMC debris least; where a description weakens a limit at all, it does
so by at most ten percent. For $\Op{1}$ the LMC-debris description reaches
three quarters of the SHM limit at $\delta = 200\,$keV and less than half
of it at $300\,$keV, while Auriga-25 reaches one half and one ninth at the same splittings. Beyond the SHM ceiling the ratio ceases to be defined,
since the SHM has no sensitivity there while the other descriptions do.

The ordering of the curves within each panel follows Table~\ref{tab:ops}.
The velocity-suppressed and doubly suppressed operators, which carry the
$\eta_1$ weighting, separate earliest and furthest from the SHM, while the
unsuppressed and the purely momentum-suppressed operators move together. At
$\delta = 300\,$keV, for instance, the LMC-debris description gives for
$\Op{7}$ and $\Op{14}$ a limit about a quarter of the SHM one, against
somewhat less than half for $\Op{1}$, $\Op{4}$ and $\Op{10}$. This is the
quantitative form of the expectation that suppressed operators are more
sensitive to the astrophysical input.

The bottom row of Fig.~\ref{fig:ratio} extends the comparison to the full
region of interest, which is the quantity relevant to a generic exclusion
limit rather than to the interpretation of the LZ event, and places the
elastic and inelastic cases side by side. Each band spans the full range of
the ratio over the three simulation-based descriptions for the operators of
one class, so that its edges mark the largest strengthening and the largest
weakening that the choice of halo can produce. The bands lie almost
entirely below unity. The simulation-based descriptions weaken a limit by
at most ten percent anywhere, whereas they strengthen it by large factors
close to the kinematic ceiling, where the SHM runs out of particles first.
The size of the effect is set by the description and modulated by the
operator within it. Underlying this is the ratio of the velocity integrals
of a given description to those of the SHM, which stays close to one
through the bulk, dips slightly below it around $600\kms$ for the
reconstruction, and then rises steeply above $700\kms$, diverging where the
SHM support ends; for Auriga-25 it exceeds one already from about $200\kms$. A limit at fixed kinematics is an efficiency-weighted average of
this function over the range of $\vmin$ that the region of interest
samples, and that range moves upward through these features as the mass
decreases or the splitting grows. For inelastic scattering at $1\,$TeV the bands stay within a few percent of unity below $50\,$keV for the FIRE-2-based descriptions, and within twenty percent for Auriga-25, open up between
$150$ and $250\,$keV, and collapse towards zero as the SHM ceiling is
approached. For elastic scattering the same structure appears as a
function of mass: the bands reach down to a few per mille at the SHM onset
near $3\,$GeV, recover to within a factor of three of unity by $7\,$GeV,
and above $20\,$GeV the FIRE-2-based descriptions agree with the SHM to better than twenty percent, above $100\,$GeV to better than five, while Auriga-25 remains ten to fifteen percent away even at large masses because of its faster bulk.

\subsection{Event-interpretation regions} \label{subsec:islands}

We now turn the question around and ask which points of the parameter space
account for the observed event together with the absence of events
elsewhere. A fit to one event alone is degenerate, because the coupling can
always be tuned to produce exactly one expected count; what breaks the
degeneracy is the rest of the data. The exclusion limits of
Sec.~\ref{subsec:method_limits} bound the coupling from above. Here we ask
instead for which $(m_\chi, \delta)$ the DM hypothesis describes
the data better than neighbouring points in the plane and, in absolute
terms, as well as any hypothesis could, and for this a likelihood is needed
rather than a counting bound.

At each point in the plane the recoil spectrum of Eq.~\eqref{eq:counts} is
smeared with the energy resolution to give an observable spectrum $f(E)$ in
events per keV at unit coupling. Three quantities characterise it: the
density at the measured energy, $f_{248} \equiv f(248\,{\rm keV})$; the
expected counts over the whole region of interest, $N_{\rm ROI}$; and the
expected counts in the high-energy sideband between $367$ and $780\,$keV,
$N_{\rm HE}$, for which we assume an acceptance $\epsilon_{\rm HE} = 0.9$
since the published efficiency is not tabulated there. With an overall
coupling strength $s \equiv (c\,m_v^{2})^{2}$ multiplying all three, the
extended unbinned likelihood of observing exactly one event, at
$248\,$keV, and nothing else is
\begin{equation}
\mathcal{L}(s) = e^{-s\,n^\prime}\,\big(s\,f_{248} + b\big),\qquad
n^\prime \equiv N_{\rm ROI} + \epsilon_{\rm HE}\,N_{\rm HE},
\label{eq:likelihood}
\end{equation}
where $b$ is the background density in the event window reported by LZ.
The first factor is the Poisson probability of no further events, the
second the probability density of the event that was seen. The coupling is
a nuisance parameter and is profiled analytically,
\begin{equation}
\hat s = \max\!\left[0,\;\frac{1}{f_{248}}\Big(\frac{f_{248}}{n^\prime}
- b\Big)\right],
\label{eq:shat}
\end{equation}
which leaves a profiled negative log-likelihood
$\Lambda(m_\chi,\delta) = \hat s\,n^\prime - \ln(\hat s\,f_{248} + b)$
that depends on the kinematics alone.

\begin{figure*}[t]
\includegraphics[width=\textwidth]{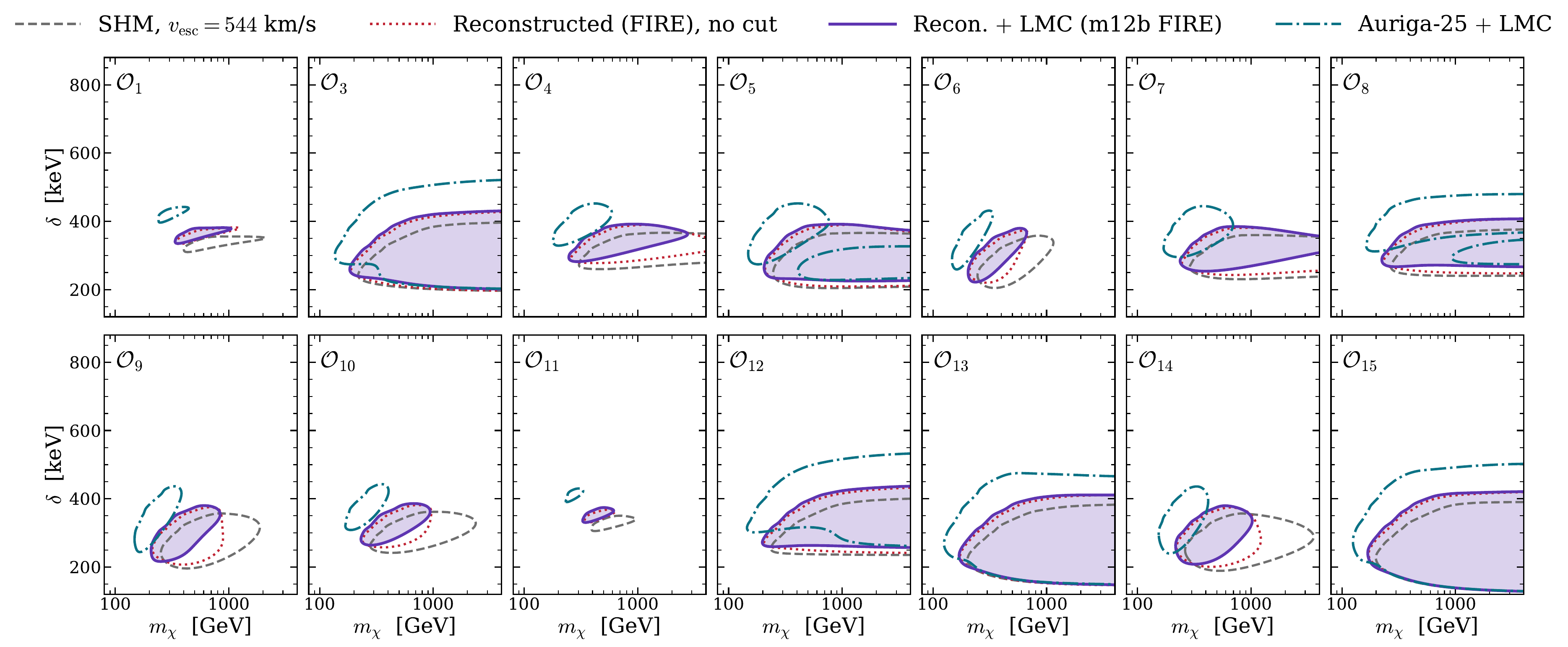}
\caption{Event-interpretation regions in $(m_\chi,\delta)$ for all fourteen
operators and each halo description: the points at which the $248\,$keV
event and the absence of other events are described acceptably,
Eqs.~\eqref{eq:qrel}--\eqref{eq:qabs}, with an isoscalar coupling. The
region for the reconstruction with the LMC debris is shaded. Regions that
remain open on the right extend beyond the $4\,$TeV edge of the scan,
since for $m_\chi \gg m_{\rm Xe}$ the kinematics, and with it the allowed
range of $\delta$, become independent of the mass.}
\label{fig:appislands}
\end{figure*}

A point is retained if it passes two tests. The first is relative,
\begin{equation}
q_{\rm rel} = 2\,[\Lambda - \Lambda_{\min}] < 4.61 ,
\label{eq:qrel}
\end{equation}
with $\Lambda_{\min}$ the minimum over the plane and $4.61$ the $90\%$ CL
threshold for two parameters. On its own this test would retain a region
even if every point in it described the event poorly, as long as they all
did so equally. The second test is therefore absolute. We construct the
spectrum that describes the data best, a single event concentrated in a
resolution-limited line at $248\,$keV with nothing elsewhere, for which
$f_{248} = 1/(\sqrt{2\pi}\,\sqrt{2}\,\sigma_E)$ and $n^\prime = 1$, and
require
\begin{equation}
q_{\rm abs} = 2\,[\Lambda - \Lambda_{\rm ideal}] < 2.71 ,
\label{eq:qabs}
\end{equation}
the one-sided $90\%$ CL threshold. This removes interpretations that place
the event at the right energy only by also predicting a broad continuum
that was not observed. The regions of Fig.~\ref{fig:appislands} are the
points that pass both tests, evaluated on a grid of $32$ masses between
$80\,$GeV and $4\,$TeV and $98$ splittings between $120$ and $900\,$keV.
The velocity distribution enters only through $f(E)$, so a change of halo
moves the regions in the way the velocity integrals of Sec.~\ref{sec:eft}
dictate: a more extended tail shifts them to larger $\delta$, a sharper
cutoff truncates them. The method is deliberately simple. With one event
there is no spectral shape to fit, and the two tests make explicit the two
ways in which an interpretation can fail, by being worse than its
neighbours and by being worse than the best description achievable.

The allowed regions are compact, as the kinematics require: inelastic
scattering confines the signal to a window around $E_R^{\star}$, and only
a narrow range of $\delta$ places that window on the event. Where the
regions sit depends on the halo. For the contact operator $\Op{1}$ the
allowed splitting runs from $304$ to $360\,$keV under the SHM, from $320$
to about $390\,$keV for the two FIRE-based descriptions, whose endpoints
differ by only about ten km/s and whose regions nearly coincide, and from $376$ to $448\,$keV for Auriga-25. The allowed masses move accordingly, from a
range reaching above $3\,$TeV under the SHM to one confined between $171$ and $468\,$GeV for Auriga-25.

The mass splitting inferred from the event is therefore uncertain by about $100\,$keV, or some thirty percent, and the mass by an order of magnitude, for reasons
that have to do with the astrophysical input alone. We stress that none of
this is evidence for DM: a single event is not a discovery, and
the elastic interpretation is excluded for every operator. What the
exercise shows is that the astrophysical input, rather than the
statistics, limits what can be inferred from a high-energy event.

\section{Conclusions}
\label{sec:conclusions}

The local DM velocity distribution becomes a leading astrophysical uncertainty whenever a DD signal probes the highest accessible velocities, yet most recent studies have relied on a single simulation realization Au-25. In this work, we instead have constructed an observationally motivated MW velocity distribution by applying the reconstruction framework developed in previous work to \textit{Gaia} DR3, and combined it with the closest LMC analogue in FIRE-2. We compared the resulting distribution with the two commonly used benchmarks, the SHM and the Au-25 MW--LMC analogue. The three descriptions agree reasonably well over the bulk of the distribution but differ substantially in their extreme tails. In particular, the FIRE-2 reconstruction extends to $835\kms$, or $846\kms$ after including the LMC-associated component, while Au-25 reaches substantially higher heliocentric speeds.

These differences become important precisely when the minimum velocity required for scattering approaches the end of the distribution. Propagating the different halo models through the complete NREFT basis for DM--nucleon scattering, we find that the kinematic reach of direct detection is presently an astrophysical systematic of order unity rather than a property of the detector alone. Across the simulation-based descriptions, the laboratory-frame endpoint spans $835$--$975\kms$, corresponding, for $m_\chi=1\,$TeV, to a maximum accessible inelastic mass splitting of $423$--$575\,$keV. Because the absolute endpoint is determined by the sparsely sampled extreme tail, a more robust comparison is given by the $99.99\%$ containment velocity, for which the corresponding ranges are $824$--$951\kms$ and $412$--$548\,$keV. Thus, limits quoted at large mass splitting or low DM mass without specifying the treatment of the high-velocity tail can differ at the factor-of-two level. Away from the kinematic ceiling, by contrast, the dependence on the halo model remains within roughly ten percent for the FIRE-2-based descriptions, and the SHM provides an adequate description; Auriga-25, whose laboratory-frame bulk is faster, departs from it by up to about twenty percent.

The high-energy event reported in the LZ extended search provides a concrete illustration. The region of $(m_\chi,\delta)$ capable of reproducing the event shifts appreciably between the SHM, the FIRE-2-based reconstruction, and Au-25 because each model supplies a different amount of phase space at the velocities required by the event. Although LMC-associated DM constitutes only $\simeq0.3\%$ of the local density in our FIRE-2 realization and is essentially invisible in the bulk of $f(v)$, it extends the laboratory-frame endpoint from $835\kms$ for the reconstructed MW component to $846\kms$, compared with $793\kms$ for the SHM, and shifts the mass splitting inferred for $\Op{1}$ by approximately $20\,$keV relative to the SHM. Once the ambient MW component has fallen away, this trace population can therefore determine the remaining kinematic reach. At the same time, the highest-velocity particles in a Solar-neighbourhood volume are represented by very small particle numbers, so these endpoint velocities should not be interpreted as sharply predicted physical cutoffs. A more robust comparison is therefore obtained from high-containment velocities, while the absolute last-particle speed should be regarded primarily as an indication of the range currently permitted by the simulations. This distinction is particularly important when quoting limits or preferred parameter regions close to the kinematic ceiling. The result is also sensitive to the Solar--LMC geometry: placing the observer in the simulated disk of the same halo yields LMC-associated particles slower than the ambient distribution they perturb. Accurately predicting this regime therefore requires dedicated simulations that both reproduce the observed MW--LMC orbital configuration and Solar-frame geometry and resolve the extreme tail with substantially larger particle statistics.

We have also examined DM--electron scattering, which becomes even more sensitive to the high-velocity tail at sufficiently low masses. We present these studies in the appendices.

Improving this situation requires both a larger statistical sample of realistic MW--LMC systems and substantially better mass resolution in the Solar neighborhood. At present, the comparison between FIRE-2 and Au-25 already shows that a single LMC analog cannot define the high-velocity tail uniquely, while the sparsity of particles at the highest speeds prevents either simulation from determining a precise endpoint. Higher-resolution zoom-in simulations, ideally spanning a broader range of MW--LMC configurations, will be needed to turn the present spread among realizations into a quantitative uncertainty. 


\begin{acknowledgments}

We thank J. Fan, P. Graham, A. Prabhu, M. Reece, N. Rodd, R. Wechsler, A. Wetzel, L. Xu, M. Escudero and K.Langhoff for helpful conversations. The work of GH was supported by the Neutrino Theory Network Fellowship with contract number 726844.
X.Z. and L.N. received support from NSF, via CAREER award AST-2337864. L.N. also acknowledges the support of the Sloan Fellowship. Large language models were used primarily for language editing and improving clarity and presentation; all scientific analysis, interpretation, and conclusions were carried out by the authors.
\end{acknowledgments}

\appendix




\section{DM-electron channel}
\label{app:electron}

DM--electron scattering provides one of the leading avenues for probing sub-GeV DM, for which nuclear recoils rapidly become kinematically inaccessible. Low-threshold searches for ionization and electronic excitations have therefore become increasingly sensitive to this mass range. DAMIC-M and PandaX-4T have now begun to reach the dark-photon benchmarks: for a relic that constitutes all of the DM, DAMIC-M excludes the freeze-in prediction above a few MeV and the PandaX-4T ionisation-only data cut into the freeze-out band, leaving sub-dominant fractions as the surviving parameter space~\citep{DAMICM:2026bench,FreezeIn:2026bm,Cheek:2025pxs}. Because
the prediction is a narrow band, whether a given model is excluded depends
on factors of order unity in the expected rate.

The astrophysical velocity distribution becomes particularly important in this channel at low DM mass. For electron scattering, the deposited energy is fixed by the electronic transition rather than directly by the momentum transfer, so the minimum incoming speed has an irreducible lower bound $\sqrt{2\Delta E/m_\chi}$ that cannot be reduced by varying $q$. As $m_\chi$ decreases, this minimum moves progressively into the high-velocity tail of $f(\mathbf{v})$, and below a few MeV the predicted rate can be supplied almost entirely by the fastest local DM particles. The low-mass reach of an electron-scattering experiment can therefore become controlled by $\vmax$ rather than by the detector threshold itself.
The NREFT framework introduced for DM--nucleon scattering in Sec.~\ref{sec:eft} can be extended to electron targets by replacing $m_N\to m_e$ and $\mathbf{S}_N\to\mathbf{S}_e$~\citep{Catena:2019gfa,Catena:2021qsr}; for semiconductor targets, the nuclear response functions are correspondingly replaced by crystal responses. Three features of this channel are particularly relevant to what follows.

First, there is no isospin: the sum over $\tau$ in Eq.~\eqref{eq:hamN}
collapses to a single set of couplings $c_j$. We factor the momentum
dependence of the mediator into the standard DM form factor
$F_{\rm DM}(q)$, taking $F_{\rm DM}=1$ for a heavy mediator and
$F_{\rm DM}=(\alpha m_e/q)^{2}$ for a light one.

Second, the electron is bounded, so the transverse velocity acquires a
dependence on the electron momentum $\mathbf{k}$,
\begin{equation}
\mathbf{v}^{\perp}_{\rm el} = \mathbf{v}
- \frac{\mathbf{k}}{m_e}
- \frac{\mathbf{q}}{2\mu_{\chi e}} ,
\label{eq:vperpel}
\end{equation}
and no longer commutes with position. It must therefore be promoted to an
operator acting on the target wavefunctions, and the rate is expressed
through a set of crystal response functions
$W_{1\text{--}5}(q,\Delta E)$~\citep{Catena:2021qsr},
\begin{equation}
\begin{split}
\frac{\mathrm{d}R}{\mathrm{d}\Delta E} \propto{}&
\sum_{i=1}^{5}\int\!\mathrm{d}q\; q\, F_{\rm DM}^{2}(q)\,
W_i(q,\Delta E)\\
&\times\Big[ A_i\,\eta_{0} + B_i\big(\eta_{1} - \vmin^{2}\eta_{0}\big)\Big],
\end{split}
\label{eq:rateE}
\end{equation}
with $A_i,B_i$ the corresponding contractions of Wilson coefficients. The
decomposition into $\eta_0$ and $\eta_1$ is unchanged; only the target
response differs, so that the same halo tables serve both channels.

Third, and most important for what follows, the kinematics are different.
Because the deposited energy $\Delta E$ is set by the electronic transition
rather than by the momentum transfer, the two are independent and
\begin{equation}
\vmin^{e} = \frac{q}{2m_{\chi}} + \frac{\Delta E}{q} ,
\label{eq:vminE}
\end{equation}
to be compared with Eq.~\eqref{eq:vminN}: the crystal absorbs momentum, so it
is the DM mass rather than the reduced mass that enters the first
term~\citep{Essig:2015cda,Catena:2021qsr}. The $\Delta E/q$ term diverges at
small $q$ and gives Eq.~\eqref{eq:vminE} a minimum at
$q_\star=\sqrt{2m_{\chi}\Delta E}$,
\begin{equation}
\vmin^{e,\star} = \sqrt{\frac{2\Delta E}{m_{\chi}}} .
\label{eq:vminstar}
\end{equation}
Unlike the nuclear case, no choice of $q$ brings $\vmin$ below this value,
and the value itself falls only as $m_\chi^{-1/2}$. For a semiconductor gap
of about an eV and a two-electron threshold, $\vmin^{e,\star}$ enters the
tail of $f(v)$ below a few MeV and exceeds the halo endpoint somewhat below
that. Sub-MeV electron recoils are therefore sourced by the fastest
particles in the halo and by nothing else, which makes this the corner of
the parameter space considered here where the astrophysical uncertainty is
largest.

\begin{figure*}[t!]
\includegraphics[width=\textwidth]{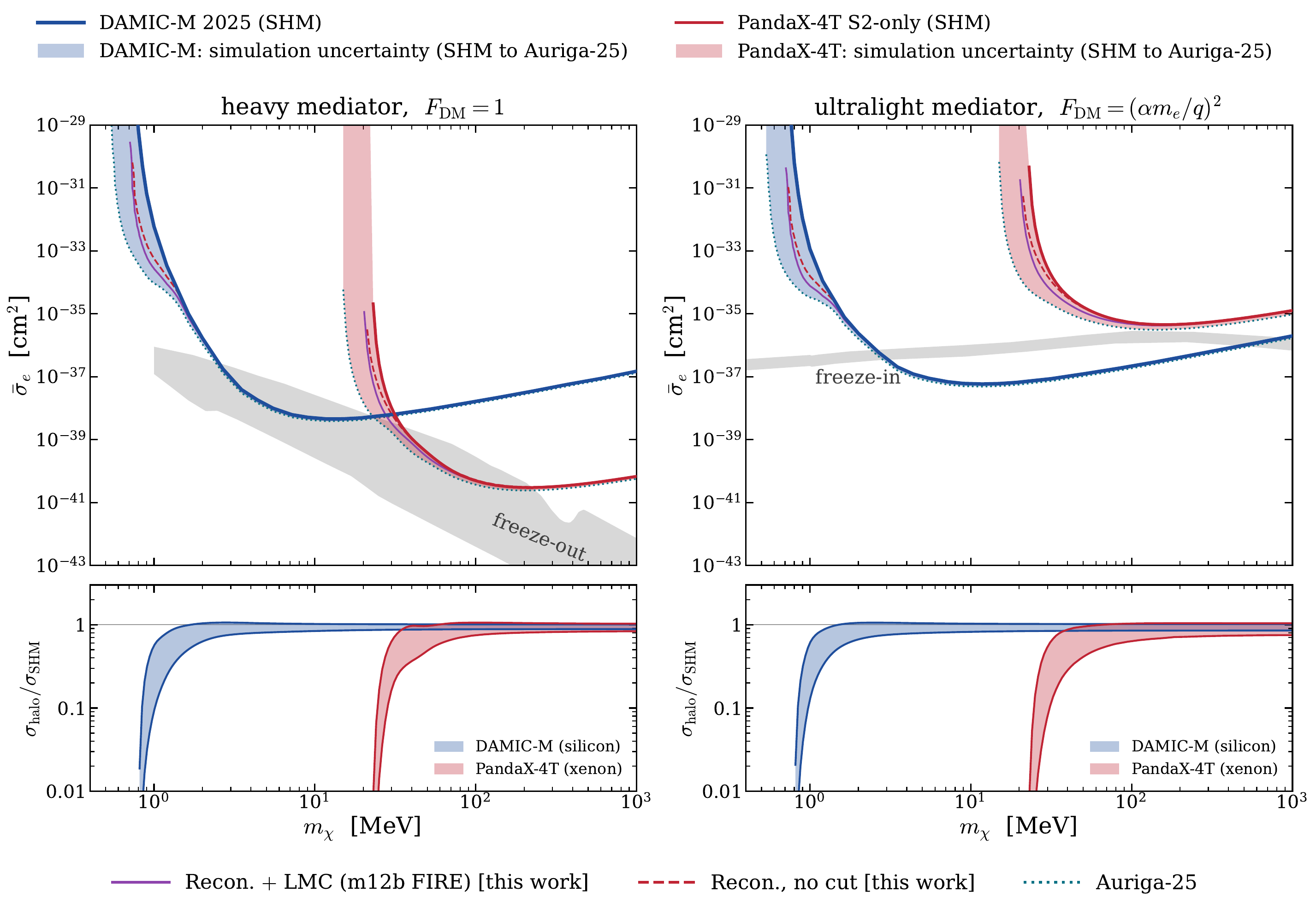}
\caption{The contact operator $\Op{1}$ in the DM--electron plane,
for a heavy mediator (left) and an ultralight one (right). Upper panels:
the DAMIC-M exclusion~\citep{DAMICM:2026bench} and the ionisation-only
($n_e = 4$--$8$) PandaX-4T limit of \citet{Cheek:2025pxs} under the SHM
(thick lines), the individual simulation-based descriptions (thin lines),
and the shaded regions spanning the simulation uncertainty from the SHM
curve to the Auriga-25 one. The grey bands are the freeze-out (left) and
freeze-in (right) relic targets, calibrated on the Snowmass
report~\citep{Essig:2022dfa}. Lower panels: the ratio
$\sigma_H/\sigma_{\rm SHM}$ as the range over the three descriptions, one
band per experiment.}
\label{fig:elo1}
\end{figure*}

\begin{figure*}[t!]
\includegraphics[width=\textwidth]{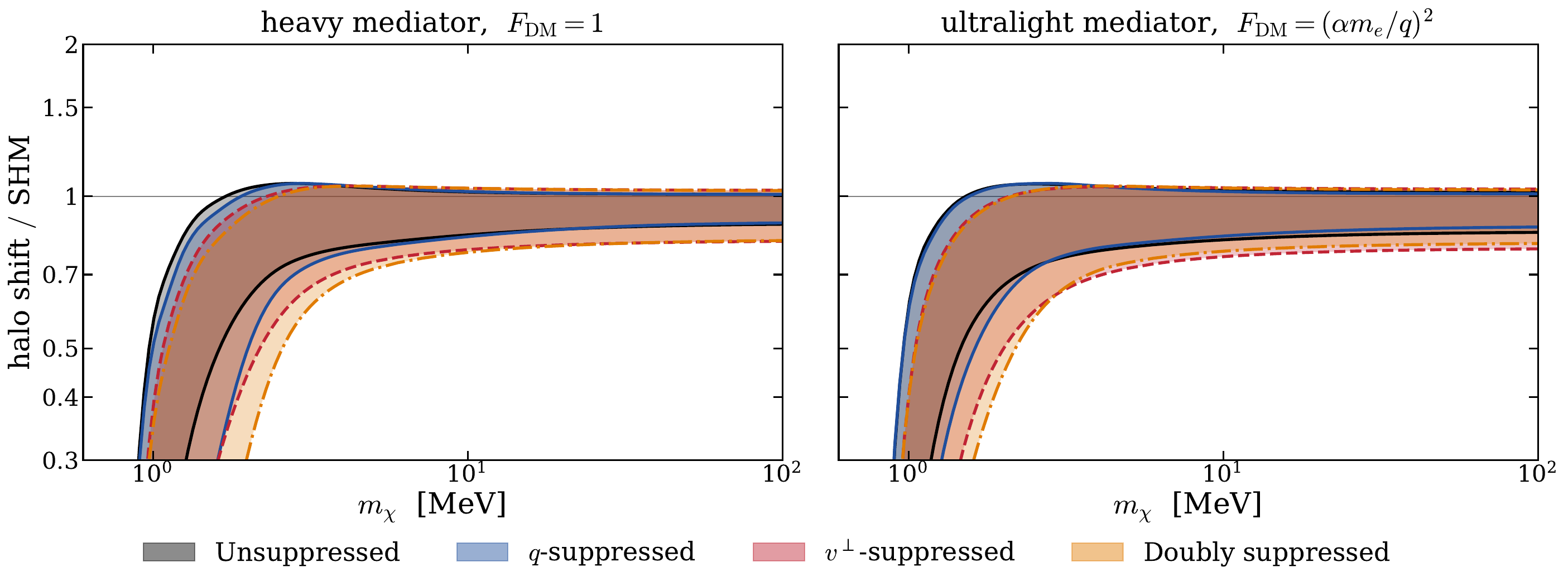}
\caption{The DM--electron analogue of the bottom row of
Fig.~\ref{fig:applimits}: the ratio of the limit obtained with a
simulation-based description to the SHM one, versus mass, for a heavy
(left) and an ultralight (right) mediator. All fourteen operators are
compressed into bands by suppression class, each band spanning the full
range over the operators in the class and the three descriptions. The
vertical axis is clipped at $0.3$.}
\label{fig:elbands}
\end{figure*}

In practice Eq.~\eqref{eq:rateE} is evaluated in the form standard for
semiconductor targets~\citep{Essig:2011nj,Essig:2015cda}. The coupling is
traded for a reference cross section at the characteristic momentum
$q_0 = \alpha m_e$,
\begin{equation}
\bar\sigma_e = \frac{\mu_{\chi e}^{2}}{16\pi m_\chi^{2} m_e^{2}}\,
\overline{\big|\mathcal{M}(q_0)\big|^{2}} ,\qquad
F_{\rm DM}(q) = \frac{\overline{|\mathcal{M}(q)|^{2}}}{\overline{|\mathcal{M}(q_0)|^{2}}} ,
\label{eq:sigmaebar}
\end{equation}
with $F_{\rm DM}$ as given above for the two limiting mediator masses. The rate
into a final-state energy $E_e$ is then
\begin{align}
\frac{dR}{d\ln E_e} &= \frac{\rho_\chi}{m_\chi}\,\frac{\bar\sigma_e\,\alpha\,m_e^{2}}
{\mu_{\chi e}^{2}\,M_{\rm cell}} \nonumber\\
&\times E_e\int\! \frac{dq}{q^{2}}\,\mathcal{I}\big(\vmin(q,E_e)\big)\,
F_{\rm DM}^{2}(q)\,\big|f_{\rm crystal}(q,E_e)\big|^{2} ,
\label{eq:rateEcrystal}
\end{align}
where $M_{\rm cell}$ is the mass of the unit cell, $|f_{\rm crystal}|^{2}$
the crystal form factor, and $\mathcal{I}$ the velocity integral, $\mathcal{I}=\eta_0$ for the
operators without explicit $v^{\perp}$ dependence and
$\mathcal{I}=\eta_1-\vmin^{2}\eta_0$ for those with it, as in the
nuclear case, Eq.~\eqref{eq:rateN}; the operators carrying powers of $q$
bring the corresponding powers of $q/m_e$ into the integrand. For silicon we
use the tabulated \textsc{QEdark} form factor~\citep{Essig:2015cda}. We do not model the conversion of $E_e$ into a number of electrons, which
depends on the ionization statistics of the detector. Instead we count
every deposition above an effective threshold $E_{\rm thr}$, and fix it
so that our SHM curve for silicon ends at the same mass as the published
DAMIC-M one. Since the lowest accessible mass is
$m_\chi=2E_{\rm thr}/\vmax^2$, this gives $E_{\rm thr}\simeq2.7\,$eV.
The value lies between the silicon gap of $1.2\,$eV and the
$5.0\,$eV needed for two electrons in a linear ionization model, as
expected when the electron yield is treated probabilistically. For xenon the atomic shell structure replaces
the crystal response; there we use \textsc{wimprates}~\citep{Aalbers:wimprates}
with our velocity integrals substituted for its own, and integrate the rate
over the $n_e = 4$--$8$ window ($E_e = 66$--$133\,$eV) of the PandaX-4T
ionisation-only analysis~\citep{Cheek:2025pxs}. What the comparison between halo descriptions depends on is where
$\vmin^\star$ meets the tail, and that is set by $E_{\rm thr}$ alone. Two
remarks apply to the results of Figs.~\ref{fig:elo1} and~\ref{fig:elbands}. First, \textsc{QEdark} tabulates the standard
response, $W_1$ in the language of \citet{Catena:2019gfa}; the four further
crystal responses probed by the $v^{\perp}$- and $q$-dependent operators are
not tabulated, and we approximate them by $W_1$ with the appropriate
kinematic weighting. This affects absolute rates operator by operator but
not the comparison between halo descriptions at fixed operator, which
depends only on the velocity integrals. Second, because $\vmin(q, E_e)$ has
a minimum at fixed $E_e$, Eq.~\eqref{eq:vminstar}, the integrand of
Eq.~\eqref{eq:rateEcrystal} vanishes identically once $\vmin^\star$ exceeds
the halo endpoint; every description therefore loses sensitivity at a
definite mass, and these masses differ between descriptions.

The same halo tables applied to a silicon target give the limits of
Fig.~\ref{fig:elo1} for $\Op{1}$ and, compressed across the operator basis,
Fig.~\ref{fig:elbands}.

We recompute the published limits of DAMIC-M~\citep{DAMICM:2026bench} and
of the PandaX-4T ionisation-only analysis~\citep{Cheek:2025pxs} for each
halo description, following the analyses of those references. For xenon we use the \textsc{wimprates} rate described above, whose native SHM our velocity integrals reproduce to about one percent; a single normalisation constant,
which absorbs exposure, efficiency and the statistical construction of the
experiment, is fixed by matching the SHM curve to the published limit, at
its minimum for the heavy mediator and at $100\,$MeV for the ultralight
one, and is then held fixed for every description. The shape of the xenon
curves is therefore entirely that of the rate, and the band between the SHM
and Auriga-25 curves is a pure velocity-integral effect. The construction
does not include the ionisation and resolution smearing of the detector,
and our SHM curve therefore terminates at the kinematic onset of the
window, slightly above the point where the published curve ends; the more
extended tails carry the reach well below that mass. For silicon, with the effective threshold fixed as above, close to the kinematic onset, where a step threshold is a poor
representation of the detector response, the curves are continued with
the shape of our own rate.

In Fig.~\ref{fig:elbands} the lower edge of each band is set by Auriga-25,
which strengthens every limit, and the upper edge by the reconstruction without the LMC debris, which weakens them slightly above a couple of MeV, where
its bulk rather than its tail is sampled. The dependence on the operator
is weak, because the effective threshold fixes the minimum speed
$\vmin^\star = \sqrt{2\Delta E/m_\chi}$ independently of the operator, so
that every class samples the same part of the tail at a given mass and the
bands lie nearly on top of one another; for the same reason the dependence
on the mediator is weak as well. The band narrows steadily with increasing
mass, from a factor of four to eight at $1.5\,$MeV to about thirty-five percent at $10\,$MeV, and it drops towards zero as the SHM rate
vanishes below $1\,$MeV. The descriptions lose sensitivity at $0.81\,$MeV
for the SHM, at about $0.72\,$MeV for the two FIRE-based descriptions and at $0.53\,$MeV for Auriga-25.

Figure~\ref{fig:elo1} also places the two experiments against the relic
targets. With the S2-only PandaX-4T analysis of \citet{Cheek:2025pxs},
whose heavy-mediator limit reaches $3\times10^{-41}\,{\rm cm}^2$, xenon
now cuts into the freeze-out band: under the SHM the limit lies inside the
band between about $35$ and $220\,$MeV, and the more extended tails open
this window at its low-mass end, down to $23\,$MeV for Auriga-25. For the
ultralight mediator the same analysis approaches the freeze-in band from
above without reaching it, coming within a factor of $1.6$ of its upper
edge under the SHM and within $1.1$ with Auriga-25, while DAMIC-M excludes
the freeze-in prediction above a mass that moves from $3.1\,$MeV under the
SHM and the FIRE-based descriptions to $2.9\,$MeV with Auriga-25. The
halo thus decides, at the level of tens of percent in mass, where the
relic targets are first reached.

Above $m_\chi \sim 10\,$MeV the halo dependence is otherwise mild, at the
level of a few to a few tens of percent, for the same reason as in the
nuclear channel: the rate samples the bulk. Below a few MeV the behaviour changes
character, because Eq.~\eqref{eq:vminstar} pushes the minimum speed into
the tail. At the lowest mass at which the SHM retains any sensitivity, the
limits obtained with the FIRE-based descriptions are stronger than the SHM
one by one to two orders of magnitude and the Auriga-25 limit by about three; the FIRE-based limits come within a factor of two of the SHM one by about $1.2\,$MeV and within a few percent by $3\,$MeV, while the Auriga-25 limit remains stronger by some twenty-five percent at $3\,$MeV. Below the SHM onset the
SHM gives no limit at all, while the FIRE-based descriptions retain
sensitivity down to about $0.7\,$MeV and Auriga-25 down to $0.53\,$MeV.

These ratios must be read with the sampling of the tail in mind. They are
largest at the masses where one description still has support at the
required speed and another does not, so their size is governed by where
each distribution is truncated rather than by a smooth systematic, and the
sampling of the last few simulation particles, together with the resolution
limit discussed in Sec.~\ref{subsec:money}, enters directly. A qualitative conclusion can still be made. In this channel the low-mass reach is set by the ratio of the detector
threshold to $\vmax^2$, $m_\chi\gtrsim2E_{\rm thr}/\vmax^2$, so that at
fixed threshold the mass at which sensitivity disappears moves by a factor of about $1.5$, and the cross section by orders of magnitude, across the
velocity distributions considered here. The effect is considerably larger
than in the nuclear channel, and it falls in the mass range that the
present generation of electron-recoil experiments is designed to probe.


\bibliography{apssamp,sections/refs_ours}

\end{document}